\documentclass[aps,prx,twocolumn,superscriptaddress,showpacs]{revtex4-2}
\usepackage{graphicx}% Include figure files
\usepackage{dcolumn}% Align table columns on decimal point
\usepackage{bm}% bold math
\usepackage{subfigure}
\usepackage{amsmath}
\usepackage{amssymb}
\usepackage{hyperref}
\usepackage{braket}
\usepackage{siunitx}
\usepackage{xcolor}
\usepackage{lipsum}
\usepackage{makecell}
\usepackage{verbatim}

\newcommand{\bk}{\boldsymbol k}

\newcommand{\bq}{\boldsymbol q}

\newcommand{\bV}{{\boldsymbol V}}

\newcommand{\ua}{\uparrow}
\newcommand{\da}{\downarrow}
\newcommand{\ra}{\rightarrow}

\newcommand{\SRO}{Sr$_2$RuO$_4$}

\begin{document}

\title{Theory of the intrinsic optical anomalous Hall responses in time-reversal symmetry breaking superconductors} 

\author{Jia-Long Zhang} 
\address{School of Physical Sciences, Great Bay University, Dongguan 523000, China}

\author{Wen Huang}
\email{huangwen@gbu.edu.cn}
\address{School of Physical Sciences, Great Bay University, Dongguan 523000, China}

\date{\today}

\begin{abstract}
Determining the symmetry of the Cooper pairing in superconductors is a central challenge in condensed matter physics. Time-reversal symmetry breaking (TRSB) is among the most distinctive symmetry-breaking pattern of the pair wavefunction. An important subset of TRSB pairing states, including the topological $p_x+ip_y$ pairing, is expected to exhibit the optical anomalous Hall effect. However, the microscopic ingredients that enable the Hall response and their underlying organizing principle in clean superconductors remain unelucidated. In this study, we develop a unified theory for TRSB multiband superconductors possessing inversion symmetry, and establish their intrinsic optical Hall response as fundamentally a quantum-geometric phenomenon. We distinguish unitary TRSB states, where TRSB originates in the orbital part of the pair wavefunction, from nonunitary states, where it is associated with spin polarization. Diagrammatically, we identify two distinct classes of microscopic processes: single-electron tunneling and Cooper-pair tunneling. Both involve at least one interband tunneling event driven by the interband velocity—an object closely tied to the interband Berry connection. We rigorously prove that interband Cooper pairing is a prerequisite for unitary TRSB states to generate intrinsic Hall response, whereas it is not absolutely needed for nonunitary states. In nonunitary states with intraband pairing alone, the Hall response can arise through single-electron tunneling processes, provided spin-orbit coupling is present. An effective electron theory obtained by integrating out the hole degrees of freedom reproduces these results and shows how pairing may endow electrons with an effective Berry curvature. Finally, we establish a symmetry selection rule for low-frequency Hall resonances: intrinsic resonances near the superconducting gap edge are allowed in nonunitary states but forbidden in unitary TRSB states. Our work thus establishes quantum geometry as a unifying principle for the optical Hall response of paired quantum matter and provides a guide for spectroscopic techniques, such as the polar Kerr effect measurement, to distinguish among different candidate TRSB pairing states in quantum materials.
\end{abstract}

\maketitle
\tableofcontents

\section{Introduction}
\label{sec:Intro}
Optical responses of superconductors reveal crucial information about the nature of the underlying Cooper pairing. The longitudinal optical conductivity, for example, can provide insights into the superconducting gap structure, the quasiparticle dynamics as well as the superfluid response~\cite{Basov2005RMP}. Furthermore, the optical anomalous Hall response---anomalous Hall effect at finite frequencies---is a powerful probe of time-reversal symmetry breaking (TRSB) in superconductors~\cite{Kapitulnik2009NJP}. Its presence also places stringent constraints on the spatial symmetry of the Cooper pairing, as not all TRSB superconducting states exhibit the Hall effect. The polar Kerr effect experiment is a direct measurement of the optical anomalous Hall conductivity~\cite{Argyres1955PR,Kapitulnik2009NJP}. Until now, this technique has been applied to several superconductors, including Sr$_2$RuO$_4$~\cite{XiaJ2006PRL}, UPt$_3$~\cite{Schemm2014Science}, Bi/Ni bilayer~\cite{Gong2017SciAdv}, and ${\mathrm{PrOs}}_{4}{\mathrm{Sb}}_{12}$~\cite{Levenson2018PRL}, providing one of the most definitive evidence of TRSB pairing in these materials. 

Drawing an analogy to the Hall effect in quantum Hall insulators~\cite{Volovik1988jetp}, the observation of Kerr rotation has often been interpreted as evidence for chiral superconductivity~\cite{Kallin2016RPP}. The prime candidates include $p_x+ip_y$ and $d_{x^2-y^2}+id_{xy}$ pairing states, where the Cooper pairs carry quantized orbital angular momentum of $\hbar$ and $2\hbar$, respectively. However, this interpretation deserves caution. In clean single-band chiral superconductors, the optical Hall effect is, in fact, forbidden by the so-called Galilean invariance, which dictates that the chiral relative motion of the paired electrons is decoupled from the Cooper pair center-of-mass (COM) motion~\cite{Read2000prb,Yip1992JLTP,Roy2008PRB,Lutchyn2008PRB}. Consequently, the Kerr effect—which probes the COM response to external electromagnetic fields—cannot detect the underlying chirality. The Galilean invariance is violated in the presence of translation symmetry breaking, such as extrinsic disorder~\cite{Goryo2008prb,Lutchyn2009prb,Konig2017prl,LiY2020prr,LiuHT2023prb} or geometric confinement~\cite{Furusaki2001prb}. These symmetry breakings readily entangle Cooper pair's relative and COM motion. 

To explain the Kerr effect observed in ultraclean superconductors, intrinsic mechanisms based on multiband effects have been suggested~\cite{Taylor2012PRL,Wysokinski2012PRL,Taylor2013JPCM,Gradhand2013PRB,Komendova2017PRL,WangZQ2017PRB,Brydon2019PRX,ZhangJL2020PRB,ZhangJL2024PRL,Huang2021PRR,Denys2021PRB,Can2021PRL,GaoYang2026prb}. Thus far, various model calculations have pointed to two distinct intrinsic pathways, distinguished by the origin of TRSB. In the first scenario, TRSB occurs in the orbital part of the Cooper pair wavefunction~\cite{Taylor2012PRL,Wysokinski2012PRL,Taylor2013JPCM,Gradhand2013PRB,Komendova2017PRL,WangZQ2017PRB,Brydon2019PRX,ZhangJL2020PRB,Can2021PRL}. The aforementioned chiral pairings belong to this category, and they are unitary states that do not exhibit spontaneous spin polarization. In the second scenario, TRSB also manifests in the spin sector, resulting in nonunitary Cooper pairing characterized by finite spin polarization~\cite{ZhangJL2024PRL,Huang2021PRR,Denys2021PRB,GaoYang2026prb,Frilen2026arXiv,footnote1,ZengM2023PRB}. 

However, no unified organizing principle has yet emerged for the intrinsic Hall response in multiband systems. The answer may lie in the several key unresolved puzzles outlined below. First, for unitary chiral superconducting states, previous model studies imply, curiously, that interband Cooper pairing is a necessary condition for the Hall response to arise~\cite{Taylor2012PRL,Wysokinski2012PRL,Brydon2019PRX,Denys2021PRB}. In contrast, it is not necessary for nonunitary states~\cite{ZhangJL2024PRL,Denys2021PRB}, where the Hall effect could emerge in a manner analogous to that in spin-orbit-coupled ferromagnets~\cite{Karplus1954PR,ZhangJL2024PRL}. To our knowledge, no rigorous derivation of this peculiar distinction between unitary and nonunitary TRSB states has appeared in the literature, and the role of SOC has only been discussed at a heuristic level~\cite{ZhangJL2024PRL}. Secondly, the microscopic processes underlying the optical Hall responses have not been thoroughly investigated. In particular, it remains unelucidated whether the Hall effect is associated with photon-excited single-electron scattering or two-electron (Cooper-pair) scattering events. Finally, although intrinsic optical excitations of Bogoliubov quasiparticles at the superconducting energy scale are allowed in clean multiband superconductors under certain conditions~\cite{Ahn2021NC}, it is unclear whether there exist optical Hall resonances in such low-frequency window. If it does, these resonances should be accessible by infrared spectroscopic techniques. As a side remark, we note that low-frequency intrinsic Hall resonances may also arise due to the so-called ``flapping" collective excitations, although this contribution to the Kerr rotation is typically vanishingly small at frequencies well above the superconducting gap~\cite{Yip1992JLTP,Yanase2026arXiv}.

In this work, we present a comprehensive theory of the intrinsic optical Hall response in clean TRSB multiband superconductors. Focusing on TRSB states that respect inversion symmetry, we fully address the above open puzzles, and demonstrate that quantum geometry~\cite{Provost1980,Torma2023PRL,LiuTY2024NSR,YuJ2025NPJ,GaoAnyuan2026RMP} serves as the fundamental underlying organizing principle. In fact, hints that quantum geometry may play a role have appeared, either implicitly or explicitly, in some previous literature. In particular, a common ingredient that recurs in some of the model studies is the interband velocity ~\cite{Taylor2012PRL,Taylor2013JPCM,ZhangJL2024PRL,Denys2021PRB,Frilen2026arXiv}, i.e., the off‑diagonal elements of the band-basis velocity matrix. These objects are intimately tied to the non-Abelian Berry connection between distinct Bloch bands, and therefore have a quantum-geometric origin. Physically, the interband velocity promotes coherent interband charge transfer~\cite{Karplus1954PR,Nagaosa2010RMP,XiaoD2010RMP}, orchestrating a violation of the aforementioned Galilean invariance in superconductors. Furthermore, considering the special circumstance of nonunitary superconductivity without interband pairing, Refs.~\onlinecite{ZhangJL2024PRL,Frilen2026arXiv} have explicitly linked one peculiar contribution to the intrinsic Hall conductivity to the spin-dependent Berry curvature of the normal-state Bloch electrons. In this study, we shall show that in more generic scenarios all of the contributions to the Hall response are inherently geometric in nature, in the sense that they all involve microscopic processes driven by the interband velocity. More broadly, the interband Berry connection and the underlying quantum geometry govern many other key aspects of the transport and electromagnetic responses of condensed matter systems~\cite{Torma2023PRL,LiuTY2024NSR,YuJ2025NPJ,GaoAnyuan2026RMP,Nagaosa2010RMP,XiaoD2010RMP}.

The remainder of the paper is organized as follows. Section \ref{sec:formalism} develops the general formalism for the optical Hall conductivity and provides a preliminary discussion of the role played by quantum geometry. Section \ref{sec:FeynmanDiagram} presents a diagrammatic analysis of the Hall conductivity in unitary and nonunitary TRSB superconducting states. By separating the diagrams into contributions associated with normal and anomalous Gor'kov Green's functions, we identify two distinct classes of microscopic processes that can drive the Hall response: single-electron tunnelings and Cooper-pair tunnelings. Both classes involve at least one interband electron tunneling, driven by the interband velocity. These analyses also allow us to elucidate the role of interband Cooper pairing and SOC. In Sec.~\ref{sec:EffectiveTheory}, we construct an effective theory that provides a qualitative understanding of the anomalous Hall effect in superconductors. Projecting out the hole degree of freedom, we obtain an effective electronic Hamiltonian that reflects the pairing-induced corrections. This effective theory reaches the same general conclusions as the diagrammatic approach. Section~\ref{sec:LowFrequency} examines the condition for the existence of low-frequency Hall resonances. Using symmetry arguments, we demonstrate that only nonunitary pairing states can support intrinsic Hall resonances near the superconducting gap edge. The paper is summarized in Sec.~\ref{sec:Summary}.

%In clean metals, momentum-conserving optical transitions can only take place between different Bloch bands. Such interband transitions are related to the nonvanishing velocity matrix elements connecting different bands, hereby referred to as interband velocity, 
%\begin{equation}
%V^{ab}_{\mu\bk} = \langle\psi_{a\bk} |\partial_{\mu}H_{\bk}|\psi_{b\bk}\rangle=(\epsilon_{a\bk}-\epsilon_{b\bk})\langle\partial_\mu \psi_{a\bk}|\psi_{b\bk}\rangle\,,
%\label{eq:BerryConn}
%\end{equation}
%with $\epsilon_{a\bk} \neq \epsilon_{b\bk}$. Here, $\epsilon_{a(b)\bk}$ and $\psi_{a(b)\bk}$ stand, respectively, for the energy dispersion Bloch wavefunction of band-$a(b)$, and $\langle\partial_\mu \psi_{a\bk}|\psi_{b\bk}\rangle$ describes an interband Berry connection. In this regard, these optical transitions are manifestations of the underlying quantum geometry of the Bloch states. 

\section{Optical Hall conductivity: formalism}
\label{sec:formalism}
Following the standard linear response theory, the charge Hall conductivity is defined as~\cite{Taylor2012PRL,Wysokinski2012PRL}
\begin{equation}
\sigma_H(\nu,\bq) = \frac{i}{2\nu}[\pi_{xy}(\nu,\bq ) - \pi_{yx}(\nu,\bq)] \,.
\label{eq:Hall}
\end{equation}
Here, $\pi_{xy}(\nu,\bq)$ is the transverse current-current correlator, which in the Matsubara-frequency representation is given by $\pi_{xy}(i\nu_m,\bq)=T/2 \Omega\sum_{\omega_n,{\bk}}\text{Tr}[\hat{\mathcal{V}}_{x\bk} \hat{G}(\bk,i\omega_n) \hat{\mathcal{V}}_{y\bk} \hat{G}(\bk,i\omega_n+i\nu_m,\bq)]$, with natural units $e=\hbar=k_B=1$. Here, $\Omega$ denotes the size of the system, $T$ is the temperature, $\omega_n$ and $\nu_m$ are the respective fermionic and bosonic Matsubara frequencies, $\hat{G}(\bk,i\omega_n) = (i\omega_n - \hat{\mathcal{H}}_{\bk})^{-1}$ denotes the Green's function associated with the Hamiltonian $\hat{\mathcal{H}}_{\bk}$, and $\hat{\mathcal{V}}_{\mu\bk}=\partial_\mu \hat{\mathcal{H}}_{\bk}$, where $\partial_\mu$ is a shorthand notation for $\partial_{k_\mu}$. In this study, we focus on the optical Hall conductivity with finite external frequency but $\bq \ra 0$, which relates to, for example, the polar Kerr effect experiment measured in the backscattering geometry~\cite{Kapitulnik2009NJP}. 

To reveal the optical transitions underlying the Hall response, it is instrumental to express the Green's function in the spectral representation
\begin{equation}
\hat{G}(\bk,i\omega_n) = \sum_{a} \frac{|\psi_{a\bk}\rangle\langle \psi_{a\bk}|}{i\omega_n-\epsilon_{a\bk}} \,,
\end{equation}
where $\hat{\mathcal{H}}_{\bk}|\psi_{a\bk}\rangle = \epsilon_{a\bk} |\psi_{a\bk}\rangle$. Substituting it into Eq.~\eqref{eq:Hall}, and after some algebra, we obtain the following result,
\begin{eqnarray}
\sigma_{H}(\nu)=\frac{i}{4\Omega \nu}\sum_{a,b,\bk}\frac{f(\epsilon_{a\bk})-f(\epsilon_{b\bk})}{ \nu+i 0^{+}+\epsilon_{b\bk}-\epsilon_{a\bk} } Q^{ab}_{xy,\bk}\,, \nonumber \\
\label{eq:AHE} 
\end{eqnarray}
with
\begin{equation}
Q^{ab}_{xy,\bk}= \langle \psi_{a\bk}|\hat{\mathcal{V}}_{x\bk}|\psi_{b\bk}\rangle\langle \psi_{b\bk}|\hat{\mathcal{V}}_{y\bk}|\psi_{a\bk}\rangle-( x \leftrightarrow y) \,.
\label{eq:Qxy}
\end{equation}
For metals or insulators, $|\psi_{a\bk}\rangle$ represents the cell-periodic wavefunction of the $a$-th Bloch band. Throughout this study, we perform the analysis in the band-basis (eigenstate) representation, in which the kinetic part of the Hamiltonian is diagonalized. Denoting the band-basis velocity matrix as $\hat{V}_{\mu\bk}$, its diagonal elements are simply the group velocities of the individual bands, $V^{aa}_{\mu\bk}= \partial_{\mu}\epsilon_{a\bk}$. Its off-diagonal elements, i.e., the interband velocities, are given by
\begin{equation}
V^{ab}_{\mu\bk}=\langle\psi_{a\bk}|\hat{\mathcal{V}}_{\mu\bk}|\psi_{b\bk}\rangle=(\epsilon_{a\bk}-\epsilon_{b\bk})\langle\partial_\mu\psi_{a\bk}|\psi_{b\bk}\rangle \,.
\end{equation}
This object contains the non-Abelian Berry connection between different bands, $i\langle\partial_\mu\psi_{a\bk}|\psi_{b\bk}\rangle$, which is deeply rooted in the quantum geometry of the Bloch electrons~\cite{Torma2023PRL,LiuTY2024NSR,YuJ2025NPJ}. 

The interband velocity depicts an interband quantum coherent transport~\cite{Karplus1954PR}, and in the present context, it also describes momentum-conserving interband optical transition upon absorption or emission of a photon polarized in the $\mu$-th direction. Hence, the sequence of virtual interband transitions described by Eq.~\eqref{eq:Qxy}, $Q^{ab}_{xy,\bk}=V^{ab}_{x\bk}V^{ba}_{y\bk}-V^{ab}_{y\bk}V^{ba}_{x\bk}$, fully characterizes the Hall response. It then follows that the characteristic resonance frequencies of the Hall conductivity match the band separation. 

In the DC limit ($\nu \ra 0$) and at zero temperature, the Hall conductivity Eq.~\eqref{eq:AHE} can be fully expressed in terms of the interband velocities as (assuming that band-$a$ is fully occupied and all other bands are empty)~\cite{Niu1985PRB}
\begin{equation}
\sigma_H = \frac{i}{2\pi\Omega}\sum_{b \neq a; \bk} \frac{V^{ab}_{x\bk} V^{ba}_{y\bk} - V^{ab}_{y\bk} V^{ba}_{x\bk}}{(\epsilon_{a\bk}-\epsilon_{b\bk})^2}  = \frac{i}{2\pi\Omega} \sum_{\bk} \mathcal{B}^a_{\bk} \,,
\label{eq:BerryCurvature}
\end{equation}
where $\mathcal{B}^a_{\bk}=i\big[  \langle\partial_x\psi_{a\bk}|\partial_y \psi_{a\bk}\rangle  -   \langle\partial_y\psi_{a\bk}|\partial_x \psi_{a\bk}\rangle \big]$ denotes the Berry curvature of band-$a$. Since the integral of the Berry curvature yields the Chern number of the band, the expression above recovers the well-established result that the DC Hall conductivity of an anomalous quantum Hall insulator is quantized and proportional to the Chern number.

The same formalism also applies to superconductors, where now $\hat{\mathcal{H}}_{\bk}$ is replaced by the corresponding Bogoliubov-de Gennes (BdG) Hamiltonian. In the Nambu particle-hole spinor basis
\begin{equation}
\hat{H}^\text{sc}_{\bk} = \begin{pmatrix}
\hat{H}_{\bk} &  \hat{\Delta}_{\bk} \\
\hat{\Delta}^{\dagger}_{\bk} & -\hat{H}^{\ast}_{\bar{\bk}}
\end{pmatrix}
\label{eq:BdG}
\end{equation}
where $\hat{H}_{\bk}$ is a diagonal matrix whose elements are given by the band energies $\epsilon_{i\bk}$. And the velocity matrix is related to the normal-state one as follows,
\begin{equation}
\hat{V}^\text{sc}_{\mu\bk}=\begin{pmatrix}
\hat{V}_{\mu\bk} & 0 \\
0 & -\hat{V}_{\mu\bar{\bk}}^\ast 
\end{pmatrix}
\label{eq:Vsc}
\end{equation}
Note that, since Cooper pairing by itself does not incur charge transport, the velocity matrix does not contain $\bk$-derivatives of the pairing potentials. As a result, optical transitions between Bogoliubov quasiparticle bands follow different selection rules than the transitions in the non-superconducting states. For example, photons cannot excite momentum-conserving transitions between the two Bogoliubov bands of a clean single-band superconductor. This is because the velocity operator is proportional to an identity matrix, i.e., $-V_{\mu\bar{\bk}}^\ast = V_{\mu\bk}$, in single-band models. 

In multiband superconductors, momentum-conserving optical transitions are possible. These processes are again facilitated by the interband velocities $V_{\mu}^{ab}$ or, the quantum geometry of the paired electrons, in a manner that depends closely on the mixed particle-hole character of the Bogoliubov quasiparticles. However, this crucial information is not sufficiently manifested in the spectral-representation formulation in Eq.~\eqref{eq:AHE}. Furthermore, multiband superconductors may exhibit both intraband and interband Cooper pairs. Microscopically, how intraband and interband Cooper pairs may act individually or in a composite manner to drive the Hall response is not clear from Eq.~\eqref{eq:AHE}. In the next section, we turn to the diagrammatic representation of Eq.~\eqref{eq:Hall}, where the desired information naturally emerges in the formulation. 

\begin{figure}
\includegraphics[width=0.8\linewidth]{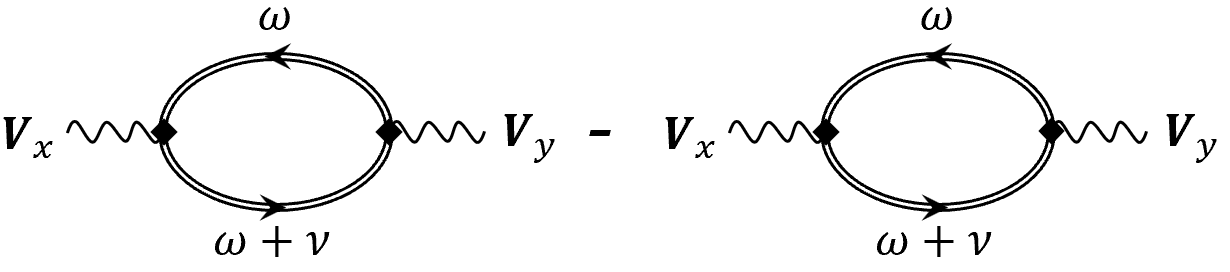}
\caption{Digrammatic representation of the intrinsic Hall conductivity within the one-loop approximation. Double lines represent the Gor'kov Green's function, while thin wavy lines at the vertex denote velocity operators.}
\label{fig:Feynman-diagram}
\end{figure}

Before proceeding to detailed analysis, it is worth emphasizing that our goal is to investigate the problem for spinful models, with SOC fully accounted for. Our study therefore will also clarify the role of SOC in the Hall response. Furthermore, unless otherwise specified, our formulation—including the pairing gap functions—will be cast entirely in the Bloch-band basis. The normal-state band structure exhibits Kramers degeneracy, $\hat{\mathcal{H}}_{\bk}|\psi_{a\bk s}\rangle = \epsilon_{a\bk}|\psi_{a\bk s} \rangle$, where $s$ indexes the two pseudospins of each band. This degeneracy complicates the definition of the pseudospins. In practice, however, one can always choose a proper basis -- for instance, the ``manifestly covariant Bloch basis" as introduced in Ref.~\onlinecite{FuL2015PRL} -- in which the Bloch wavefunction $|\psi_{a\bk s}\rangle$ for each pseudospin species evolves smoothly across the momentum space. For notational simplicity, pseudospin will henceforth be referred to simply as spin in most circumstances. Adding the spin indices, the interband velocity can be defined as
\begin{equation}
V^{ab,ss^\prime}_{\mu\bk}=\langle\psi_{a\bk s}|\hat{\mathcal{V}}_{x\bk}|\psi_{b\bk s^\prime}\rangle=(\epsilon_{a\bk}-\epsilon_{b\bk})\langle\partial_\mu\psi_{a\bk s}|\psi_{b\bk s^\prime}\rangle \,.
\label{eq:Vband}
\end{equation}
Evidently, velocity matrix elements with $s\neq s^\prime$ but $a=b$ shall vanish. However, due to the non-Abelian nature of the interband Berry connecton, there can generally be finite interband velocity between states of the same spin ($s = s^\prime$) and between states of opposite spins ($s \neq s^\prime$). By Hermiticity, normal-state time-reversal (TRS) and inversion symmetries, these velocity elements satisfy the following respective relations,
\begin{eqnarray}
V^{ab,ss^\prime}_{\mu\bk} &=&(V^{ba,s^\prime s}_{\mu \bk})^\ast  \label{eq:hermiticity}\\
V^{ab,ss^\prime}_{\mu\bk} &=&-(V^{ab,\bar{s}\bar{s^\prime}}_{\mu\bar{\bk}})^\ast \label{eq:TRS} \\
V^{ab,ss^\prime}_{\mu\bk} &=& -V^{ab,s s^\prime}_{\mu\bar{\bk}} \label{eq:inversion} \,,
\end{eqnarray}
where $\bar{s}=-s$ and $\bar{s^\prime}=-s^\prime$. The above relations also combine to give
\begin{equation}
V^{ab,ss^\prime}_{\mu\bk} = (V^{ab,\bar{s}\bar{s^\prime}}_{\mu\bk})^\ast = V_{\mu\bk}^{ba,\bar{s^\prime}\bar{s}} \,.
\end{equation}
These equalities will be used frequently in the diagrammatic analyses. 

We will focus on time-reversal-symmetry-breaking (TRSB) quasi-two-dimensional superconducting order that preserves spatial inversion symmetry, without restricting the analysis to any particular lattice geometry. To support a Hall effect within the $xy$-plane, such systems must also break all vertical-plane mirror symmetries~\cite{Cho2016PRL}. This additional constraint forbids the Hall response in certain TRSB states, such as the $s+id_{x^2-y^2}$ and $d_{x^2-y^2}+ig_{xy(x^2-y^2)}$ pairings in crystals with $D_{4h}$ symmetry---to which \SRO~belongs. In the present study, we study TRSB unitary and nonunitary pairings that meet the above symmetry criteria, while a more systematic group-theoretic symmetry analysis will be presented in a separate study. 

An important clarification is needed at this point. Although the concept of nonunitary pairing has appeared in broader contexts~\cite{RamiresJPCM2022}, we refer to it exclusively as Cooper pairing that breaks TRS in the pseudospin part of the pair wavefunction and one that generally exhibits nonvanishing net spin polarization. These states typically feature different gap amplitudes for the pairing between spin-up electrons and that between spin-down electrons, i.e., $|\Delta_{ab}^{ss}| \neq |\Delta_{ab}^{\bar{s}\bar{s}}|$. On account of the inversion symmetry, this condition naturally constrains nonunitary pairings to be spin-triplet in the pseudospin representation. For example, the $D_{4h}$ group hosts three distinct nonunitary states: an $E_u$ state with a three‑dimensional $d$‑vector that includes both $(k_x,k_y)\hat{z}$ and $k_z(\hat{x},\hat{y})$ components~\cite{Huang2018PRL}, and two so‑called mixed helical $p$‑wave states, $A_{1u}+iA_{2u}$ and $B_{1u}+iB_{2u}$~\cite{Huang2021PRR,Denys2021PRB,ZhangJL2024PRL,GaoYang2026prb}. By contrast, unitary pairing satisfies $\Delta^{s\bar{s}} = \eta \Delta^{\bar{s}s}$, with $\eta=1$ for spin‑triplet and $\eta=-1$ for spin‑singlet pairing; here TRS is broken by the orbital part of the Cooper pair wavefunction, i.e., $\Delta^{s\bar{s}}_{\bk} \neq (\Delta^{s\bar{s}}_{\bk})^\ast$. Spin-singlet sates have the additional constraint that $\Delta^{ss}=0$.

\section{Diagrammatic analyses and microscopic interpretation}
\label{sec:FeynmanDiagram}

The one-loop Feynman diagram associated with Eq.~\eqref{eq:Hall} is shown in Fig.~\ref{fig:Feynman-diagram}. Here, the vertices represent velocity operators, and the double lines represent the full Gorkov Green's function. Decomposing the Green's function into normal and anomalous parts returns two classes of diagrams (A and B) as given in Table~\ref{tab:diagram_scattering}. Their corresponding contributions to the Hall response are derived in this section. In all diagrams the band indices $a\neq b$. For any superconducting state, diagrams with $a=b$ can be independently shown to have vanishing contribution to the Hall conductivity, consistent with the general understanding that multiband superconductivity must be involved to generate intrinsic Hall effect. These analyses shall elucidate not only the role of quantum geometry, but also the underlying microscopic processes that drive the intrinsic Hall reponse. We shall see that the two classes of diagrams correspond, respectively, to two distinct types of processes: single-electron tunnelings and Cooper-pair tunneling. These processes actively involve the tunneling of at least one electron across different bands, facilitated by interband velocity (or the interband Berry connection). In passing, we note that analogous microscopic processes have also been shown to be responsible for the quantum-geometric superfluid weight in generic multiband superconductors~\cite{HuYJ2025prb}. Our analysis also reveals a key distinction: nonunitary superconducting states may exhibit intrinsic Hall responses in the absence of interband Cooper pairing, while unitary TRSB states necessarily require it. 

\begin{table}[]
    \centering
    \begin{tabular}{c}
\includegraphics[width=\linewidth]{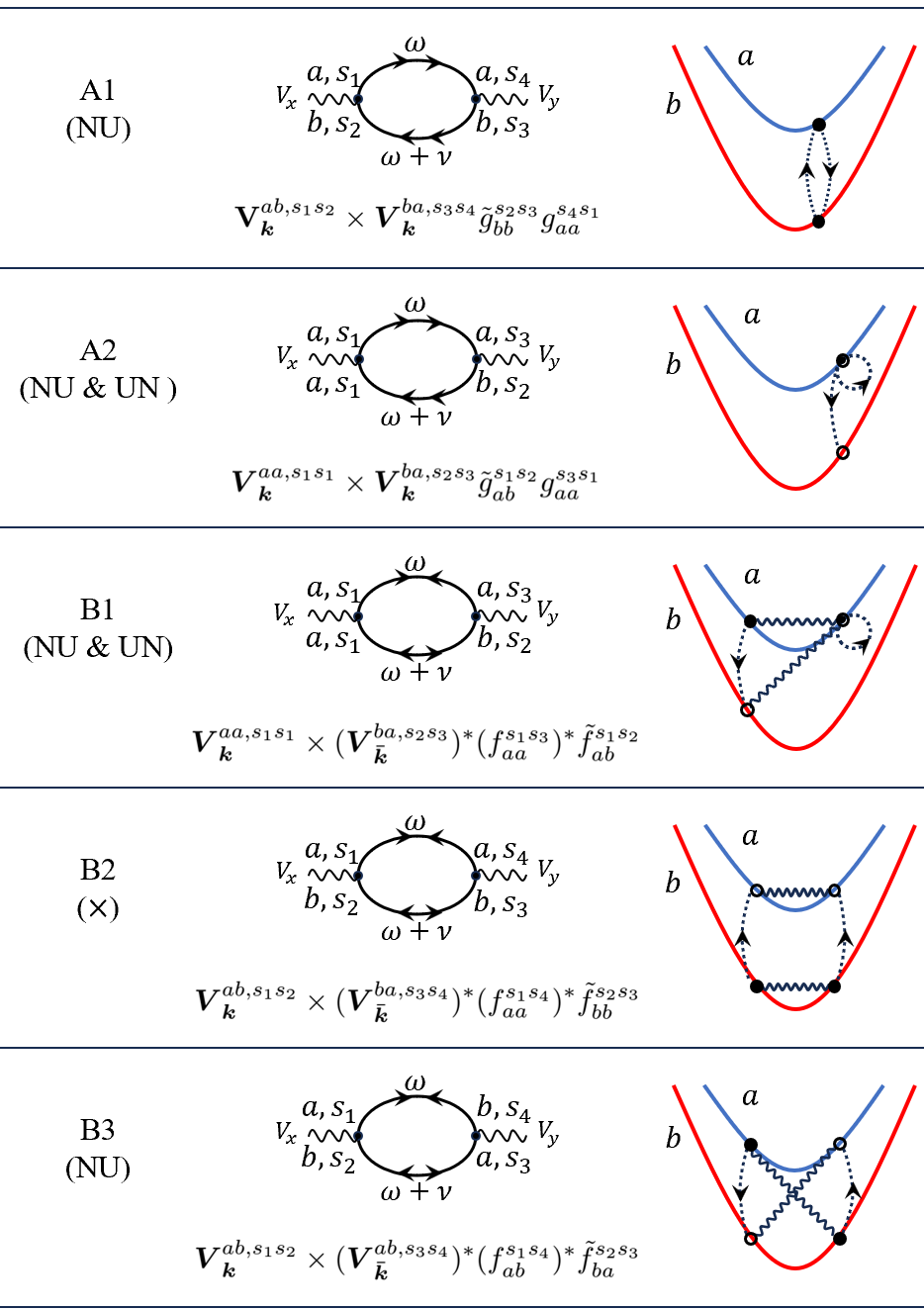}
    \end{tabular}
\caption{Feynman diagrams decomposed from the full one-loop diagram of the anomalous Hall conductivity in Fig.~\ref{fig:Feynman-diagram} and the corresponding microscopic processes. Second column: Representatives of the decomposed Feynman diagrams. The counter diagrams with velocity vertices interchanged, which also carry an additional minus sign, are not shown here. These diagrams are labeled A1, B2, B1, B2, and B3; and the TRSB superconducting states in which they could have finite contributions are indicated in the parentheses, with `NU' and `UN' standing for nonunitary and unitary TRSB states, respectively. In the present discussion of superconducting states with inversion symmetry, the B2 contribution vanishes, as indicated by the ``$\times$" symbol in the respective parentheses. The integrands that appear in the calculation of the Hall conductivity are given below each diagram, where $g$ ($\tilde{g}$) and $f^\ast$ ($\tilde{f}$) denote the normal and anomalous parts of the Gorkov Green's function, $a$ and $b$ label the bands, and $s_p$ with $p=1,2,3,4$ label the spins. The velocity cross product in each integrand involves at least one interband velocity. Third column: Schematic illustration of the microscopic tunneling processes underlying the corresponding diagrams in the second column. Here, the blue and red curves represent normal-state band dispersion, and the thick wavy lines denote Cooper pairing between the connected electrons. The arrows on the dotted lines indicate the direction of the tunneling driven by the intraband or interband velocity. The two classes of processes are distinguished by whether the interband tunneling is of a single electron (Class A) or a Cooper pair (Class B). The two fermion lines at the left vertices of the A2 and B1 diagrams have the same spin label $s_1$, because intraband velocity does not have finite element associated with opposite spins. Other choices of $\{s_p \}$ in each diagram depend on the pairing state considered.}
    \label{tab:diagram_scattering}
\end{table}

\subsection{Class-A diagrams}
The first type of diagrams (Class A in Table ~\ref{tab:diagram_scattering}) contain only the normal parts of the Gorkov Green's function, $g^{ss^\prime}_{ab} (i\omega_n,\bk)$. The two velocity vertices are both given by the electron part of the velocity operator, $V_{x\bk}$ and $V_{y\bk}$. The surviving Class A diagrams, with appropriate arrangement of the band and spin indices, are shown in Table~\ref{tab:diagram_scattering} A1 and A2.

{\it A1 diagrams}: In close inspection, at each wavevector $\bk$ the A1 diagrams return the following finite contributing terms to the Hall conductivity (omitting, for simplicity, constants unimportant for the purpose of the present discussion.),  
\begin{eqnarray}
& & \sum_{a\neq b,\{s_p\}} V^{ab,s_1s_2}_{x\bk}\tilde{g}^{s_2 s_3}_{bb} V^{ba,s_3 s_4}_{y\bk} g^{s_4s_1}_{aa} - (x \leftrightarrow y) \nonumber \\
%&=& \sum_{a\neq b,s,s^\prime} (V^{ab,ss^\prime}_{x\bk} V^{ba,s^\prime s}_{y\bk} - V^{ab,ss^\prime}_{y\bk} V^{ba,s^\prime s}_{x\bk})  \tilde{g}^{s^\prime s^\prime}_{bb}g^{ss}_{aa} \, . \nonumber \\
& =& \sum_{a\neq b,\{s_p\}} \bV^{ab,s_1 s_2}_{\bk} \times \bV^{ba,s_3 s_4}_{\bk} \tilde{g}^{s_2s_3}_{bb}g^{s_4s_1}_{aa} \,.
\label{eq:TypeA1}
\end{eqnarray}
Here, the symbol ``$\{s_p\}$'' under the summation sign with $p=1,2,3,4$ represents summation over each spin label, $\bV^{ab,s_p s_q}_{\bk}=(V^{ab,s_q s_q}_{x\bk}, V^{ab,s_p s_q}_{y\bk})$, and we have adopted the shorthand notation $g^{s_4s_1}_{aa}\equiv g^{s_4s_1}_{aa}(i\omega_n,\bk)$ and $\tilde{g}^{s_2 s_3}_{bb}\equiv \tilde{g}^{s_2 s_3}_{bb}(i\nu_n+i\omega_n,\bk)$. Note that the cross product vanishes if $a=b$. The remaining cross product of the interband velocities is related to the interband Berry curvature. Microscopically, this contribution can be understood to originate from virtual back-and-forth interband tunneling of a single electron at wavevector $\bk$, facilitated by orthogonal interband velocities as illustrated in Table~\ref{tab:diagram_scattering}. 

As we prove in detail in App.~\ref{app:A0}, the Class A1 contribution to the intrinsic Hall response vanishes in unitary superconducting states. To see how this contribution can be finite in nonunitary states, let us focus on Eq.~\eqref{eq:TypeA1} with $s_1=s_4$ and $s_2=s_3$, under which scenario the velocity wedge product is purely imaginary. Further utilizing the relations Eqs.~\eqref{eq:hermiticity}-\eqref{eq:inversion}, one obtains
\begin{equation}
\bV^{ab,s_1s_2}_{\bk} \times \bV^{ba,s_2 s_1}_{\bk} = -\bV^{ab,\bar{s}_1\bar{s}_2}_{\bk} \times \bV^{ba,\bar{s}_2 \bar{s}_1}_{\bk} \,.
\label{eq:spinBC}
\end{equation}
Hence, to ensure nonvanishing Eq.~\eqref{eq:TypeA1}, the normal Green's function must satisfy $\tilde{g}^{s_2 s_2}_{bb}g^{s_1s_1}_{aa} \neq \tilde{g}^{\bar{s}_2 \bar{s}_2}_{bb}g_{aa}^{\bar{s}_1\bar{s}_1} $, which is indeed the case in nonunitary superconducting states where $\Delta^{ss} \neq \Delta^{\bar{s}\bar{s}}$. For example, let us take a simplified nonunitary model with $\Delta^{s\bar{s}}=\Delta^{\bar{s}s}=0$ and assume only intraband pairing, it follows that
\begin{equation}
g^{ss}_{aa} (\omega_n,\bk) = \frac{i\omega_n -\epsilon_{a\bk}}{-w_n^2 + \epsilon_{a\bk}^2 + |\Delta^{ss}_{aa,\bk}|^2} \,. 
\end{equation}
Then it is readily seen that $g^{ss}_{aa} \neq g^{\bar{s}\bar{s}}_{aa}$. The conclusion does not change when interband pairing is also considered. 

The above conclusion explicitly relies on having nonvanishing spin-dependent Berry curvature $\bV^{ab,s_1s_2}_{\bk} \times \bV^{ba,s_2 s_1}_{\bk}$, which switches sign when the spin indices change to their opposite (see Eq.~\eqref{eq:spinBC}). This implies a normal-state spin Hall effect, which is present only when there exists finite SOC. Hence, our conclusion can also be interpreted from a more intuitive perspective as follows. In the metallic phase of any inversion-symmetric spin–orbit-coupled model~\cite{Karplus1954PR,XiaoD2010RMP,Nagaosa2010RMP}, the spin Hall effect is generally present, although the charge Hall effect is absent due to exactly opposite contributions from the two spin species. In nonunitary superconducting states, however, this perfect cancellation breaks down, leading to a residual charge Hall conductivity. The same conclusion was obtained in Ref.~\onlinecite{ZhangJL2024PRL}, where the Hall conductivity was found to depend linearly on the pairing-induced spin polarization. Hence, the Class A1 mechanism is essentially similar to what underlies the Hall effect in (anomalous) Hall metals or insulators. For completeness, in App.~\ref{app:A} we present a more complete derivation of the Hall conductivity associated with the A1 diagrams. 

{\it A2 diagrams}:~The A2 diagrams in Table~\ref{tab:diagram_scattering} correspond to
\begin{equation}
\sum_{a\neq b,\{s_p\}} \bV^{aa,s_1s_1}_{\bk}\times \bV^{ba,s_2s_3}_{\bk}\tilde{g}_{ab}^{s_1s_2}g_{aa}^{s_3s_1} \,.
\label{eq:TypeA2}
\end{equation}
In distinction to Eq.~\eqref{eq:TypeA1}, this expression involves both intraband and interband velocities. Here, one electron on band-$a$ is first transferred to itself via intraband velocity and then tunneled to band-$b$ by virtue of the interband velocity, as illustrated in Table~\ref{tab:diagram_scattering}. A necessary condition for this contribution to be finite is that the superconducting state has finite interband pairing, because the interband element of the Green's function $g_{ab}$ would otherwise vanish. This contribution may appear in both unitary and nonunitary TRSB phases. Appendix~\ref{app:B} derives the Hall conductivity for a minimal spinless two-band model of a unitary chiral superconductor and explicitly isolates this contribution. The derivation can be readily generalized to spinful unitary and nonunitary TRSB pairing states. 

\subsection{Class-B diagrams}
The second class of diagrams contains only the anomalous parts of the Gor'kov Green's function, $f^{ss^\prime}_{ab}(i\omega_n,\bk)$. In this study, the anomalous Green's functions are defined according to the following convention: $f_{ab}^{s_{1}s_{2}} \ra \langle c_{a\bk s_{1}}^{\dagger} c_{b\bar{\bk} s_{2}}^{\dagger}\rangle$ and $(f_{ab}^{s_{1}s_{2}})^{*}\ra \langle c_{a\bar{\bk} s_{1}} c_{b\bk s_{2}} \rangle $. Here, nonzero elements of \(f^{ss^\prime}_{ab}\) with \(a=b\) and \(a\neq b\) are closely tied to the presence of intraband and interband Cooper pairings, respectively. Shown in Table~\ref{tab:diagram_scattering} are the three representative diagrams, B1, B2 and B3, classified according to the type of microscopic tunneling processes involved. In distinction with the Class A diagrams, here one velocity vertex is electron-like, $V_{\mu \bk}$; while the other is hole-like, $-(V_{\nu \bar{\bk}})^\ast$. Hence, unlike the two single-electron tunnelings occurring at the same wavevector in the class-A diagrams, the class-B processes correspond to the tunneling of particles at opposite wavevectors --- equivalent to the transfer of Cooper pairs, as will be elaborated below.

{\it B1 diagrams}:~Since $V^{aa,s \bar{s}}_{\mu\bk}=0$, the B1 diagrams is expressed as follows,
\begin{eqnarray}
    &&\sum_{a\neq b,\{s_p\}}\bV^{aa,s_1s_1}_{\bk} \times (\bV_{\bar{\bk}}^{ba,s_2s_3})^\ast (f^{s_1s_3}_{aa})^\ast \tilde{f}^{s_1s_2}_{ab} \nonumber \\
    &+& \sum_{a\neq b,\{s_p\}} \bV^{ba,s_2s_3}_{\bk} \times (\bV_{\bar{\bk}}^{aa,s_1s_1})^\ast 
    (f^{s_2 s_1}_{ba})^\ast \tilde{f}^{s_3 s_1}_{aa}
    \nonumber \\
    &=&\sum_{a\neq b,\{s_p\}}\bV^{aa,s_1s_1}_{\bk}\times \left[ \bV^{ba,s_2s_3}_{\bk} (f^{s_2s_1}_{ba})^\ast \tilde{f}^{s_3s_1}_{aa}  \right. \nonumber \\
   & &-\left. \bV^{ab,s_3s_2}_{\bk} (f^{s_1s_3}_{aa})^\ast \tilde{f}^{s_1s_2}_{ab} \right]\,,  
    \label{eq:B1AHC}
\end{eqnarray} 
observing the shorthand notation $f^{s_1s_1}_{aa}\equiv f^{s_1s_1}_{aa}(i\omega_n,\bk)$ and $\tilde{f}^{s_2s_3}_{ba}\equiv \tilde{f}^{s_2s_3}_{ba}(i\nu_n+i\omega_n,\bk)$ (the same in the subsequent analysis). In the last line, we have used the hermiticity and inversion symmetry relations, i.e., Eqns.~\eqref{eq:hermiticity} and \eqref{eq:inversion}. 

To obtain a physical description of the underlying microscopic processes, it is important to note that, in the original operator representation $V_{\mu\bk}^{aa,s_1 s_2} \ra V_{\mu\bk}^{aa,s_1s_2} c^\dagger_{a\bk s_1}c_{a\bk s_2}$, and $(V^{ba,s_2s_3}_{\mu\bar{\bk}})^\ast \ra (V^{ba,s_2s_3}_{\mu\bar{\bk}})^\ast c_{b\bar{\bk}s_2} c^\dagger_{a\bar{\bk}s_3}$. Hence, the B1 diagrams can be interpreted as the tunneling at wavevector $\bk$ of a spin-$s_1$ electron in band-$a$ to itself, joined by the tunneling at wavevector $-\bk$ of a spin-$s_3$ hole in band-$a$ to a spin-$s_2$ hole state in band-$b$. Since the latter is equivalent to the tunneling of an electron in the opposite direction, in combination these processes can be viewed as the transfer of an interband Cooper pair to an intraband pair in band-$a$, i.e., $c_{b\bar{\bk}s_2} c_{a\bk s_1} \ra c_{a\bar{\bk}s_3} c_{a\bk s_1}$, as sketched in Table~\ref{tab:diagram_scattering} (Class B1). As can be seen, this Cooper-pair transfer is facilitated by the joint action of intraband and interband velocities, $V^{aa}$ and $V^{ab}$. 

This contribution can be finite in both unitary and nonunitary TRSB states. In App.~\ref{app:B}, the Hall conductivity associated with the B1 diagrams is derived for a spinless unitary pairing model. Together with terms originating from the A2 diagrams, i.e., Eq.~\eqref{eq:TypeA2}, they constitute the full Hall response of a unitary TRSB state, as shown in App.~\ref{app:B}.

The B1 diagrams involve the scattering between intraband and interband Cooper pairs. In what follows, we separately discuss two other distinct scenarios: one where the transfer occurs solely between intraband pairs, and the other where it occurs solely between interband pairs, namely, the B2 and B3 diagrams in Table~\ref{tab:diagram_scattering}.

{\it B2 diagrams}:~The B2 diagrams involve only the transfer of intraband pairs, and is described by the following expression,
\begin{eqnarray}
&& \sum_{a\neq b}\sum_{\{s_p\}}  \boldsymbol{V}^{ab,s_1s_2}_{\bk} \times (\boldsymbol{V}^{ba,s_3s_4}_{\bar{\bk}})^\ast (f^{s_1s_4}_{aa})^\ast  \tilde{f}^{s_2s_3}_{bb}  \nonumber \\ 
&=& - \sum_{a\neq b}\sum_{\{s_p\}}  \boldsymbol{V}^{ab,s_1s_2}_{\bk} \times \boldsymbol{V}^{ab,s_4s_3}_{\bk} (f^{s_1s_4}_{aa})^\ast  \tilde{f}^{s_2s_3}_{bb}  \,,
\label{eq:B2}
\end{eqnarray}
Following the interpretation in association with the B1 diagrams, the underlying physical processes here can be viewed as the interband-velocity-driven transfer of intraband Cooper pair from band-$b$ to band-$a$, i.e., $c_{b\bar{\bk}s^{\prime\prime\prime}} c_{b\bk s^\prime} \ra c_{a\bar{\bk}s^{\prime\prime}} c_{a\bk s}$, as sketched in Table~\ref{tab:diagram_scattering} (Class B2). However, this contribution vanishes for both unitary and nonunitary TRSB superconducting states, as we show below.  

For unitary spin-singlet states, because the pairing takes place only between opposite spins, the spin indices in the B2 diagrams must satisfy $s_4=\bar{s}_1$ and $s_3=\bar{s}_2$. Hence Eq.~\eqref{eq:B2} becomes,
\begin{eqnarray}
-\sum_{a\neq b}\sum_{s_1 s_2}  \boldsymbol{V}^{ab,s_1s_2}_{\bk} \times \boldsymbol{V}^{ab,\bar{s}_1\bar{s}_2}_{\bk} (f^{s_1\bar{s}_1}_{aa})^\ast  \tilde{f}^{s_2\bar{s}_2}_{bb}  \,.
\label{eq:B2singlet}
\end{eqnarray}
Using the following relation for singlet states $(f_{aa}^{s_1\bar{s}_1})^\ast \tilde{f}_{bb}^{s_2 \bar{s}_2} = (f_{aa}^{\bar{s}_1s_1})^\ast \tilde{f}_{bb}^{\bar{s}_2 s_2}$, it can be shown that \eqref{eq:B2singlet} vanishes after summation over the spin indices. In other words, in a unitary spin-singlet pairing state, the B2 diagrams (i.e., the interband tunneling of intraband Cooper pairs) do not contribute to the Hall response.

For spin-triplet states, unitary or nonunitary, the general contribution follows as,
\begin{eqnarray}
&&-\sum_{a\neq b}\sum_{s_1,s_2}  \boldsymbol{V}^{ab,s_1s_2}_{\bk} \times \boldsymbol{V}^{ab,s_1s_2}_{\bk} (f^{s_1 s_1}_{aa})^\ast  \tilde{f}^{s_2s_2}_{bb} \nonumber \\
&& -\sum_{a\neq b}\sum_{s_1,s_2}  \boldsymbol{V}^{ab,s_1s_2}_{\bk} \times \boldsymbol{V}^{ab,\bar{s}_1\bar{s}_2}_{\bk} (f^{s_1\bar{s}_1}_{aa})^\ast  \tilde{f}^{s_2\bar{s}_2}_{bb} \nonumber \\
&& -\sum_{a\neq b}\sum_{s_1,s_2}  \boldsymbol{V}^{ab,s_1s_2}_{\bk} \times \boldsymbol{V}^{ab,s_1\bar{s}_2}_{\bk} (f^{s_1 s_1}_{aa})^\ast  \tilde{f}^{s_2\bar{s}_2}_{bb}    \,.
\label{eq:B2triplet}
\end{eqnarray}
Here, the first (second) line originates from interband transfer between the $S_z= \pm 1$ ($S_z=0$) intraband spin-triplet Cooper pairs, while the third line is a consequence of interband tunneling between the $S_z=\pm 1$ and $S_z=0$ intraband pairs. It is apparent that the first line vanishes, and the next two lines also yield a null result upon summation over spins and on account of the relation $f^{s\bar{s}}_{aa}=f^{\bar{s}s}_{aa}$ for triplet pairing. Therefore, the interband transfer of intraband triplet pairs similarly does not generate Hall response. 

The null result associated with the B2 diagrams holds in general, regardless of whether interband pairing is present. The above findings also explain why unitary TRSB states with only intraband pairing do not produce intrinsic Hall effect.

{\it B3 diagrams}:~Following the interpretation in the above, the B3 diagrams describe the tunneling between interband pairs, e.g., from $\Delta^{s\bar{s}}_{ab}$ to $\Delta^{s\bar{s}}_{ba}$, 
\begin{eqnarray}
&& \sum_{a\neq b}\sum_{\{s_p\}}  \boldsymbol{V}^{ab,s_1s_2}_{\bk} \times (\boldsymbol{V}^{ab,s_3s_4}_{\bar{\bk}})^\ast (f^{s_1s_4}_{ab})^\ast  \tilde{f}^{s_2s_3}_{ba}  \nonumber \\ 
&=& - \sum_{a\neq b}\sum_{\{s_p\}}  \boldsymbol{V}^{ab,s_1s_2}_{\bk} \times \boldsymbol{V}^{ba,s_4s_3}_{\bk} (f^{s_1s_4}_{ab})^\ast  \tilde{f}^{s_2s_3}_{ba}  \,.
\label{eq:B3}
\end{eqnarray}
According to the same reasoning as in the analysis for B2 diagrams, this contribution is identically zero for unitary TRSB states, independent of the existence of intraband pairing. However, it is straightforward to check that it can be finite for nonunitary states. 

Concluding this section, we saw that the Hall response can be attributed to disctinct classes of single-electron or Cooper-pair transfer processes. These processes are governed by the same geometric object---the interband Berry connection, in the regard that they must involve interband tunneling of at least one electron --- driven by interband velocity, i.e., the interband Berry connection between the Bloch bands. This establishes quantum geometry as the unifying principle that governs the intrinsic Hall response. However, we stress that the geometric effect does not always manifest in the form of the quantum geometric tensor~\cite{Provost1980} of the individual Bloch bands. For example, in Eq.~\eqref{eq:B1AHC} associated with the B1 diagrams, only one interband velocity is involved, thus the velocity cross products cannot be expressed in terms of the quantum metric or Berry curvature of the Bloch wavefunctions. 

Finally, we also saw that interband pairing must be present for the intrinsic Hall effect to arise in unitary TRSB states, the reason behind which will become much clearer in the next section. In contrast, the intrinsic Hall response in nonunitary states can arise through the A1 processes that do not absolutely require the presence of interband pairing, although there must exist finite SOC for this mechanism to take effect.

\section{Effective electronic theory}
\label{sec:EffectiveTheory}
The above diagrammatic analyses offer a physically intuitive description of the physical processes underlying the intrinsic Hall response. In this section, we provide an alternative perspective by investigating the Hall response of an effective electron-only Hamiltonian that incorporates the pairing-induced corrections. This is achieved by projecting out the hole sector of the BdG Hamiltonian Eq.~\eqref{eq:BdG}~\cite{GaoYang2026prb}, 
\begin{equation}
\hat{H}_{\bk} \ra \hat{H}^\prime_{\bk} =\hat{H}_{\bk} + \hat{\Delta}^\dagger_{\bk}(\hat{H}_{\bk}+\hat{H}^\ast_{\bar{\bk}})^{-1}\hat{\Delta}_{\bk} \,.
\end{equation}
The electron part of the BdG Hamiltonian reads
\begin{equation}
\hat{H}_{\bk} = \begin{pmatrix}
\epsilon_{a\bk}  &    ~  \\
~&  \epsilon_{b\bk} & ~ \\
~& ~  & \ddots
\end{pmatrix} \otimes s_0
\end{equation}
Here, $s_0$ is the identity matrix that operates in spin subspace. The Hall response of this effective theory can then be analyzed in conjunction with the normal-state velocity matrix $\hat{V}_{\mu\bk}$. In the following, we illustrate the power of this theory for the case of unitary pairings, while a separate derivation for nonunitary pairing is presented in App.~\ref{app:EETnonunitary}. 

We consider the example of a two-band unitary odd-parity superconductor, while noting that the formalism can be equally applied to even-parity superconducting states. Take a case with both intraband and interband pairings,
\begin{equation}
\hat{\Delta}_{\bk} = \begin{pmatrix}
\Delta_{aa,\bk}  &  \Delta_{ab,\bk} \\
\Delta_{ba,\bk} &  \Delta_{bb,\bk} 
\end{pmatrix}\otimes s_{x} \,,
\end{equation}
the effective electronic Hamiltonian then follows as follows:
\begin{equation}
\hat{H}^\prime_{\bk}=\begin{pmatrix}
\epsilon_{a\bk}^\prime  &  \lambda_{\bk}  \\
\lambda_{\bk}^\ast &  \epsilon_{b\bk}^\prime
\end{pmatrix}\otimes s_0 \,,
\label{eq:effHelectron}
\end{equation}
with 
\begin{eqnarray}
\lambda_{\bk} &=& \frac{\Delta_{aa,\bk}^\ast\Delta_{ab,\bk}}{2\epsilon_{a\bk}}+\frac{\Delta_{ba,\bk}^\ast\Delta_{bb,\bk}}{2\epsilon_{b\bk}} \,, \nonumber \\
\epsilon_{a\bk}^\prime &= & \epsilon_{a\bk}+\frac{|\Delta_{aa,\bk}|^2}{2\epsilon_{a\bk}}+\frac{|\Delta_{ab,\bk}|^2}{2\epsilon_{b\bk}} \,, \nonumber \\
\epsilon^\prime_{b\bk}&=&\epsilon_{b\bk}+\frac{|\Delta_{bb,\bk}|^2}{2\epsilon_{b\bk}}+\frac{|\Delta_{ab,\bk}|^2}{2\epsilon_{a\bk}} \,.
\end{eqnarray}
While the above Hamiltonian Eq.~\eqref{eq:effHelectron} is ill-defined at Fermi wavevectors where the pairing cannot be treated as a perturbation, this effective theory qualitatively captures the essential feature of the pairing-induced corrections at other general wavevectors. As one can see, if interband pairing is absent, i.e., if $\Delta_{ab,\bk}=0$, the correction appears merely as wavevector-dependent energy shift in the diagonal elements. This alone is not sufficient to generate intrinsic Hall response in unitary TRSB states. On the other hand, interband pairing joins with intraband pairing to induce an effective interband hybridization taking the form of $\lambda_{\bk}$. After lengthy algebra, the transverse current correlator in the Matsubara-frequency representation, $\pi_{xy}(i\nu_n,\bq=0 ) - \pi_{yx}(i\nu_n,\bq=0)$, can be shown to be given by the following integrand summed over $\omega_n$ and $\bk$, as well as over the spin index,
\begin{equation}
\frac{i \text{Im}[\lambda_{\bk}(\boldsymbol{V}_{\bk}^{aa,ss}-\boldsymbol{V}_{\bk}^{bb,ss})\times\boldsymbol{V}_{\bk}^{ab,ss}] }{2[(i\omega_n-\frac{\epsilon^\prime_{a\bk}+\epsilon^\prime_{b\bk}}{2})^2-E_{\bk}^2 ] [(i\omega_n+i\nu_m-\frac{\epsilon^\prime_{a\bk}+\epsilon^\prime_{b\bk}}{2})^2-E_{\bk}^2 ]}\,.
\label{eq:AHCeffectiveTheory}
\end{equation}
Where $E_{\bk}=\sqrt{(\frac{\epsilon_{a\bk}^{\prime}-\epsilon_{b\bk}^{\prime}}{2})^2+|\lambda_{\bk}^2|}$. This expression agrees qualitatively with the exact result in App.~\ref{app:B}. In particular, consistent with our diagrammatic analyses, the form of the numerator makes it clear that both intraband and interband pairings must be involved, and that the interband velocity also plays an indispensable role. All of these ingredients are necessary for the paired electrons to acquire effective Berry curvature that is ultimately responsible for the intrinsic Hall response. 

To see how Eq.~\eqref{eq:AHCeffectiveTheory} may produce finite Hall conductivity, it is important to note that the interband pairing $\Delta_{ab,\bk}$ does not necessarily share the same form as the intraband pairings $\Delta_{aa,\bk}$ and $\Delta_{bb,\bk}$. In fact, the symmetry of the interband pairing is jointly determined by the form of $\Delta_{ab,\bk}$ and the symmetry of the two bands~\cite{ChenWP2021PRR}. For example, for a model in the $D_{4h}$ group consisting of spin-orbit-coupled $d_{xz}$ and $d_{yz}$ orbitals on each site of a square lattice~\cite{ChenWP2021PRR}, the two bands at the $\Gamma$-point have total angular momentum $j_z = \pm 3/2$ and $j_z = \pm 1/2$, respectively. In the case of unitary chiral $p$-wave Cooper pairing with total angular momentum $\hbar$, the intraband pairing is simply given by $\Delta_{ii,\bk} \propto k_x+ik_y$, while the interband pairing formed between the $j_z=\pm 3/2$ and $j_z=\mp 1/2$ states shall take the form $\Delta_{ab,\bk} \propto k_x-ik_y$. As a consequence, the induced interband hybridization has the form $\lambda_{\bk} \sim (k_x-ik_y)^2$. One can then readily see that Eq.~\eqref{eq:effHelectron} effectively describes the Hamiltonian of an anomalous Hall metal.

\section{Low-frequency Hall resonances}
\label{sec:LowFrequency}
Having established the general theory of the intrinsic Hall response, we now turn to a more subtle issue about the resonance frequency of the Hall conductivity. Traditional Kerr effect experiments use photons with energy of the order of one electrovolt, far exceeding the scale of the typical millielectrovolt superconducting gap. Hence, they are sensitive to Hall resonances occurring at electrovolt frequency scales. One natural question is, can the intrinsic Hall resonances appear at far-infrared frequencies comparable to $2\Delta$? In fact, a similar question has been asked about the longitudinal optical conductivity, for which the condition for the appearance of low-frequency resonances turns out to be less restrictive~\cite{Ahn2021NC}. Microscopically, such low-frequency responses would originate from optical transitions between the low-lying Bogoliubov quasiparticle bands separated by the superconducting gap, which can be probed by terahertz or microwave spectroscopic techniques~\cite{Armitage2508,BrounD_prb2025,Broun_arXiv2025}. Through symmetry analysis, we show in this section that such a low-frequency intrinsic Hall effect could emerge in nonunitary pairing states but not in unitary states. 

\begin{figure}
\includegraphics[width=8cm]{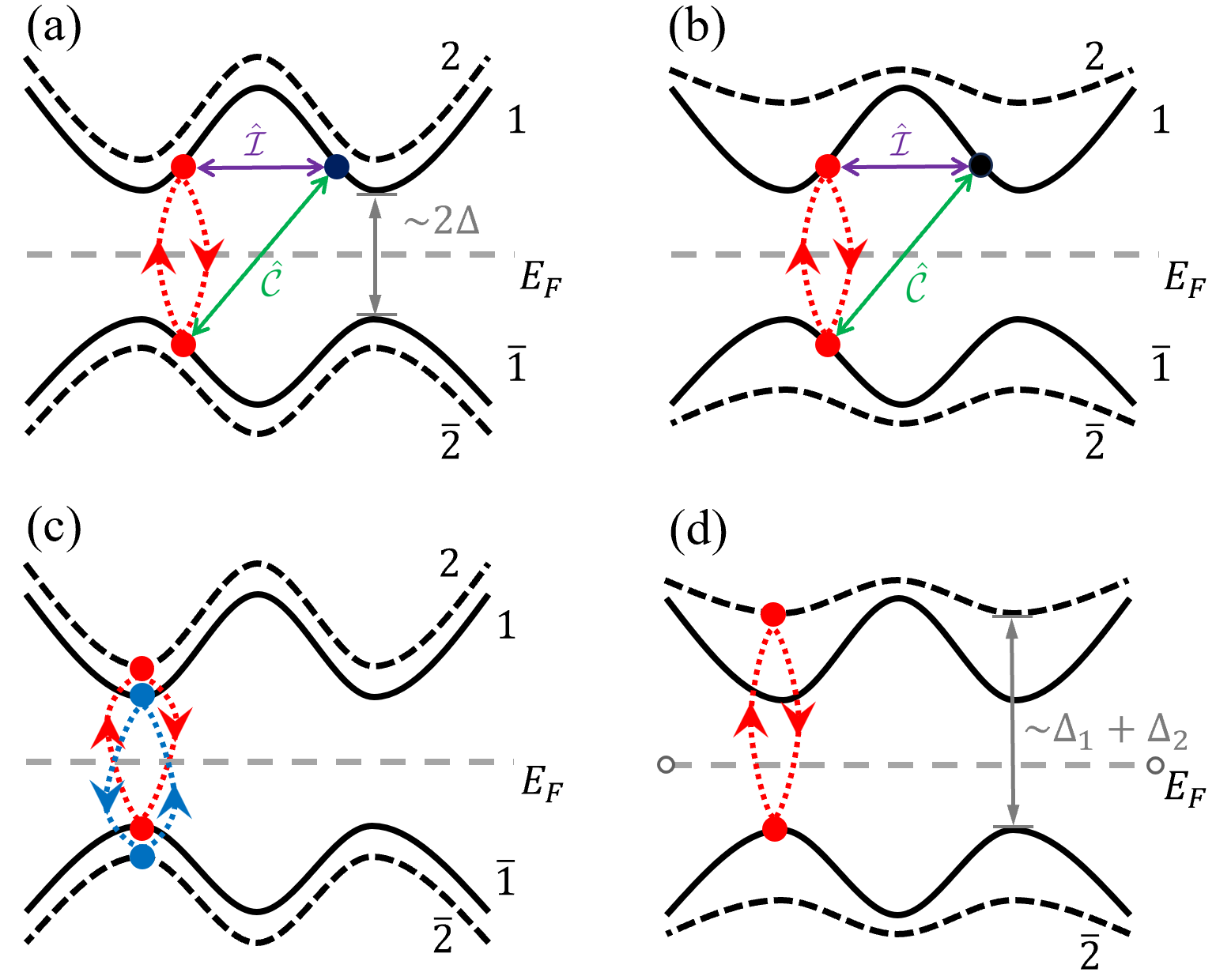}
\caption{Graphic representation of the virtual optical transitions between low-energy Bogoliubov quasiparticles for analyzing the low-frequency optical Hall responses. The left panel [(a) and (c)] is for unitary pairing states, where the degenerate Bogoliubov bands are represented by slightly displaced solid and dashed curves. The right panel [(b) and (d)] is for nonunitary states. The Bogoliubov bands related by the particle-hole symmetry are labeled by $i$ and $\bar{i}$, with $i=1,2$. The green and purple double arrows in (a) and (b) connect the pair of quasiparticle states related by particle-hole ($\hat{\mathcal{C}}$) and inversion ($\hat{\mathcal{I}}$) symmetries, respectively. The red and blue dotted curves with arrows indicate the optical transitions between Bogoliubov quasiparticle states. The Hall response originates from the virtual back-and-forth transitions between any occupied and unoccupied pair of states, and these two transitions are driven by orthogonal velocity components. The vertical gray double arrows in (a) and (d) indicate the energy gap between low-lying Bogoliubov bands.}
\label{fig:sketch}
\end{figure}

Let us first consider the particle-hole ({\it ph}) symmetry in association with the operator $\hat{\mathcal{C}}=\tau_x\mathcal{K}$, where $\tau_x$ operates in the Nambu {\it ph} subspace and $\mathcal{K}$ denotes complex conjugation operation. Specifically,
\begin{equation}
\hat{\mathcal{C}}^{-1}\hat{H}^\text{sc}_{\bk}\hat{\mathcal{C}}=-\hat{H}^\text{sc}_{\bar{\bk}} 
\end{equation}
Note that $\hat{\Delta}_{\bk}^\dagger=-\hat{\Delta}_{\bar{\bk}}$ for both even and odd parity pairings. Under the {\it ph} operation, the velocity operator becomes
\begin{equation}
\hat{\mathcal{C}}^{-1}\hat{V}^\text{sc}_{\mu\bk}\hat{\mathcal{C}}= \begin{pmatrix} 
-\hat{V}_{\mu\bar{\bk}} &  0 \\
0  & \hat{V}_{\mu\bk}^\ast 
\end{pmatrix} = -\hat{V}^\text{sc}_{\mu\bar{\bk}} \,.
\end{equation}
Inversion symmetry is implemented through
\begin{equation}
\hat{\mathcal{I}}^{-1}\hat{H}^\text{sc}_{\bk}\hat{\mathcal{I}}=\hat{H}^\text{sc}_{\bar{\bk}} \,.
\end{equation}
Here, the inversion operator reads
\begin{equation}
\hat{\mathcal{I}}= \begin{pmatrix}
\hat{I}  & 0  \\
0 & \eta \hat{I} 
\end{pmatrix} \,,
\end{equation}
where $\eta=1$ and $-1$ for even and odd parity pairings, respectively, and the operator $\hat{I}$ transforms the normal-state Hamiltonian and the pairing matrix as follows, $\hat{I}^{-1} \hat{H}_{\bk} \hat{I}= \hat{H}_{\bar{\bk}}$ and $\hat{I}^{-1} \hat{\Delta}_{\bk} \hat{I}= \hat{\Delta}_{\bar{\bk}}$. Furthermore, the inversion operation takes the velocity operator to
\begin{equation}
\hat{\mathcal{I}}^{-1}\hat{V}^\text{sc}_{\mu\bk}\hat{\mathcal{I}}=\hat{V}^\text{sc}_{\mu\bar{\bk}} \,.
\end{equation}
For a state given by $\hat{H}^\text{sc}_{\bk}|\phi_{n\bk}\rangle = E_{n\bk}|\phi_{n\bk}\rangle$, its {\it ph} counterpart is
$|\phi_{\bar{n}\bar{\bk}}\rangle=\hat{\mathcal{C}}|\phi_{n\bk}\rangle$, which satisfies $\hat{H}^\text{sc}_{\bk}|\phi_{\bar{n}\bar{\bk}}\rangle = E_{\bar{n}\bar{\bk}}|\phi_{\bar{n}\bar{\bk}}\rangle= -E_{n\bk}|\phi_{\bar{n}\bar{\bk}}\rangle$. Here, we have used $n=1,2,....$ and $\bar{n}=-n$ to index the Bogoliubov bands related by the {\it ph} symmetry (see Fig.~\ref{fig:sketch}). By inversion symmetry, $|\phi_{\bar{n}\bk}\rangle=\hat{\mathcal{I}}|\phi_{\bar{n}\bar{\bk}}\rangle$, with $\hat{H}^\text{sc}_{\bk}|\phi_{\bar{n}\bk}\rangle =E_{\bar{n}\bk} |\phi_{\bar{n}\bk}\rangle=-E_{\bar{n}\bar{\bk}} |\phi_{\bar{n}\bk}\rangle$. One therefore has $E_{\bar{n}\bk}= -E_{n\bk}$ due to the joint {\it ph} and inversion symmetries associated with the operator $\hat{\Gamma}=\hat{\mathcal{C}}\hat{\mathcal{I}}$, i.e., $\hat{\Gamma} |\phi_{n\bk}\rangle = |\phi_{\bar{n}\bk}\rangle$. Pairs of states related by the above-discussed symmetries exist in both unitary and nonunitary pairings, as depicted in Fig.~\ref{fig:sketch} (a) and (b). 

Now consider the superconducting version of the transition processes Eq.~\eqref{eq:Qxy} taking place between occupied and unoccupied Bogoliubov bands, 
\begin{eqnarray}
\langle \phi_{\bar{n}\bk}|\hat{V}^\text{sc}_{x\bk}|\phi_{m\bk}\rangle\langle \phi_{m\bk}|\hat{V}^\text{sc}_{y\bk}|\phi_{\bar{n}\bk}\rangle-( x \leftrightarrow y) \,.
\label{eq:Qxysc}
\end{eqnarray}
Inserting the identity operator $\hat{\Gamma}^{-1}\hat{\Gamma}$ to both sides of the velocity operators, the above expression can be shown to become
\begin{eqnarray}
\langle \phi_{n\bk}|(-\hat{V}^\text{sc}_{x\bk})|\phi_{\bar{m}\bk}\rangle\langle \phi_{\bar{m}\bk}|(-\hat{V}^\text{sc}_{y\bk})|\phi_{n\bk}\rangle-( x \leftrightarrow y)\,.
\end{eqnarray}
This expression is exactly the opposite of Eq.~\eqref{eq:Qxysc} when $m=n$. Hence, the corresponding optical transitions do not contribute to the Hall response. Representative transitions between states in the lowest-lying Bogoliubov bands around the Fermi level, $|\phi_{1}\rangle$ and $|\phi_{\bar{1}}\rangle$, are displayed in Fig.~\ref{fig:sketch} (a) and (b) for unitary and nonunitary states, respectively. 

In the case of unitary pairing, the Bogoliubov bands are doubly degenerate. We now address whether optical transitions between pairs of lowest-energy states not related by $\hat{\Gamma}$ can generate Hall response. As an example, the Bogoliubov bands labeled 1 and 2 in Fig.~\ref{fig:sketch} (c) satisfy $E_{1\bk} = E_{2\bk}$, and their corresponding quasiparticle wavefunctions are related by $ |\phi_{2\bk}\rangle = \hat{\gamma} |\phi_{1\bk}\rangle$, with
\begin{equation}
   \hat{\gamma} = \begin{pmatrix}
   \hat{s}_x & 0 \\
   0 & \eta\hat{s}_x
   \end{pmatrix} \,.
\end{equation}
Here, $\hat{s}_x$ is the $x$-component of the Pauli matrix that operates in the pseudospin subspace, which flips the pseudospins. Note that $\hat{\gamma}=\hat{\gamma}^{-1}$. Due to the normal-state TRS and inversion symmetry, the velocity operator satisfies
\begin{equation}
    \hat{\gamma} \hat{V}^{\text{sc}}_{\mu\bk}\hat{\gamma}^{-1} = (\hat{V}^{\text{sc}}_{\mu\bk})^\ast  \,.
\end{equation}
We consider the following two distinct sets of transition processes (also sketched Fig.~\ref{fig:sketch} (c)),
\begin{eqnarray}
&& \langle \phi_{\bar{1}\bk}|\hat{V}^\text{sc}_{x\bk}|\phi_{2\bk}\rangle\langle \phi_{2\bk}|\hat{V}^\text{sc}_{y\bk}|\phi_{\bar{1}\bk}\rangle-( x \leftrightarrow y) \nonumber \\
&+& \langle \phi_{\bar{2}\bk}|\hat{V}^\text{sc}_{x\bk}|\phi_{1\bk}\rangle\langle \phi_{1\bk}|\hat{V}^\text{sc}_{y\bk}|\phi_{\bar{2}\bk}\rangle-( x \leftrightarrow y) \,.
\label{eq:QxyscDoubleDegenerate}
\end{eqnarray}
Inserting the identity operator $\hat{\gamma}^{-1}\hat{\gamma}$ into both sides of the velocity operators in the second line, the above expression can be shown to become,
\begin{eqnarray}
&& \langle \phi_{\bar{1}\bk}|\hat{V}^\text{sc}_{x\bk}|\phi_{2\bk}\rangle\langle \phi_{2\bk}|\hat{V}^\text{sc}_{y\bk}|\phi_{\bar{1}\bk}\rangle-( x \leftrightarrow y) \nonumber \\
&+& \langle \phi_{\bar{1}\bk}|(\hat{V}^\text{sc}_{x\bk})^\ast|\phi_{2\bk}\rangle\langle \phi_{2\bk}|(\hat{V}^\text{sc}_{y\bk})^\ast|\phi_{\bar{1}\bk}\rangle-( x \leftrightarrow y) \,.
\label{eq:QxyscDoubleDegenerate1}
\end{eqnarray}
It is then easy to prove that the two lines cancel exactly. Hence, the type of transitions depicted in Fig.~\ref{fig:sketch} (c) cannot generate net Hall conductivity. In summary, optical transitions between the lowest-lying Bogoliubov bands in unitary superconductors do not produce Hall response. The resonance of the Hall conductivity occurs at frequency scale of the order of the normal-state interband separation, which is typically much higher than $2\Delta$. 

\begin{figure}
\includegraphics[width=7.5cm]{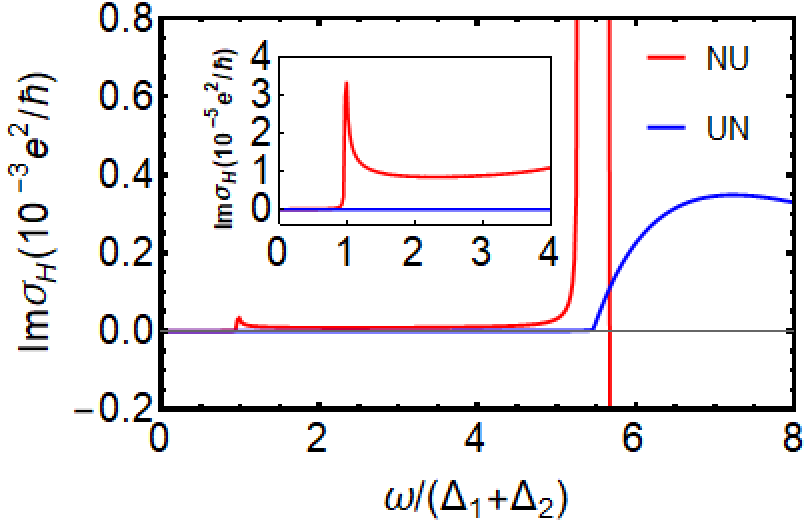}
\caption{Frequency dependence of the imaginary part of the Hall conductivity as a function of $\omega/(\Delta_1+\Delta_2)$. Details of the model calculation are given in App.~\ref{app:C}. The red and blue curves correspond to the nonunitary (NU) and unitary (UN) superconducting states, respectively. The inset shows an enlarged view of the low-frequency regime of the Hall conductivity. A resonance peak emerges near $\omega \approx \Delta_1+\Delta_2$ in the nonunitary state, where $\Delta_1$ and $\Delta_2$ denote the gap size of the Bogoliubov bands as defined in App.~\ref{app:C}.}
\label{fig:low w Hall}
\end{figure}

In nonunitary states, the Bogoliubov bands are split (Fig.~\ref{fig:sketch} (b) and (d)), and the low-lying quasiparticle states $|\phi_{1\bk}\rangle$ and $|\phi_{2\bk}\rangle$ are no longer related by the $\hat{\gamma}$ symmetry. Hence, the transition processes described by Eq.~\eqref{eq:QxyscDoubleDegenerate}, as shown in Fig.~\ref{fig:sketch} (d), can generate Hall response in generic spin-orbit-coupled models. This contribution is associated with resonance frequency around $\omega=\Delta_1+\Delta_2$, where $2\Delta_i~(i=1,2)$ denotes the energy gap between the $i$-th and $\bar{i}$-th Bogoliubov bands. This is numerically verified using a two-band model on a square lattice, as given in App.~\ref{app:C}. Figure \ref{fig:low w Hall} shows the representative Hall conductivity obtained from the calculation for a unitary chiral $p$-wave and a nonunitary $p$-wave pairing. As is evident, the nonunitary state does exhibit a low-frequency Hall resonance near the gap edge, with a cutoff frequency approximately equal to $\Delta_1+\Delta_2$. In contrast, the cutoff frequency of the unitary state lies much beyond the gap edge.

It is worth noting that, disorder scattering can also induce Hall response at low frequencies around $\omega =2\Delta$~\cite{Goryo2008prb,Lutchyn2009prb,Konig2017prl,LiY2020prr,LiuHT2023prb}. In practice, this extrinsic contribution can be distinguished from the intrinsic one through Kerr effect measurements on samples with controlled disorder. Finally, our analysis also indicates that, if inversion symmetry is broken, either in the normal state or in the superconducting state through mixed-parity pairings, the low-frequency Hall responses can be allowed.

\section{Summary and outlook}
\label{sec:Summary}
In this work, we carried out a comprehensive theoretical study of the intrinsic optical Hall effect in TRSB multiband superconductors and established quantum geometry as the unifying organizing principle behind it. First, adopting a diagrammatic approach, we identified two distinct classes of microscopic processes that can give rise to intrinsic anomalous Hall responses: (i) tunneling of single electron, and (ii) tunneling of Cooper pairs. These processes involve at least one interband electron transfer driven by interband velocity, i.e., the interband Berry connection. In this regard, the intrinsic Hall response is intimately tied to the quantum geometry of the paired electrons (not that of the Bogoliubov quasiparticles). Furthermore, we rigorously showed that interband pairing is a prerequisite for the intrinsic anomalous Hall effect to occur in unitary TRSB superconductors, but not in nonunitary superconductors. In the latter, finite SOC is essential to promote intrinsic charge Hall effect in the absence of intraband pairing. 

In an alternative perspective, we developed an effective theory that examines how the superconducting pairing modifies the electron degrees of freedom, where the Hall response can be analyzed from a resultant effective electronic Hamiltonian. The key conclusions derived from this theory agree fully with those of the diagrammatic analysis. We further proved that intrinsic low-frequency optical Hall resonances near the superconducting gap edge may exist in nonunitary pairing states but are absent in unitary chiral states. 

Our work carries special significance for the experimental efforts to identify the nature of the pairing in TRSB superconductors, such as the still highly controversial \SRO~\cite{Kallin2016RPP,Mackenzie2017npjQM,Wysokinski2019CM,LiuY2021JSNM,Huang2021CPB,Maeno2024JPSJ,Maeno2024NP}. The polar Kerr effect operates under stringent symmetry constraints, allowing it to differentiate between different subsets of candidate TRSB pairing states. Until now, Kerr effect measurements have been performed at optical frequencies considerably exceeding the superconducting gap size~\cite{XiaJ2006PRL,Schemm2014Science,Gong2017SciAdv,Levenson2018PRL}.  Our study suggests that Kerr rotation at low frequencies near the gap edge in pristine samples can further distinguish between pairings in the subset of Kerr-sensitive states. 

In the future, our work can be naturally extended to systems lacking inversion symmetry, where the symmetry breaking originates either in the normal state or from a parity-mixed Cooper pairing~\cite{GaoYang2026prb,HuLH2021PRB}. A further unaddressed question concerns the relation between the optical Hall response and the orbital magnetization~\cite{ZhuJ2026arXiv,WangZhi2026arXiv}. In particular, for nonunitary states, it remains to be investigated whether or how the Hall response may be simultaneously influenced by the orbital and spin magnetization. These will be left for future study.

\section*{Acknowledgements} We acknowledge helpful discussions with Wen Sun and Zhiqiang Wang. This work is supported by the NSFC under Grant No.~12374042, Guangdong Provincial Quantum Science Strategic Initiative (GDZX2601007), the Guangdong Science and Technology Department under Grant 2022A1515011948, and the Dongguan Key Laboratory of Artificial Intelligence Design for Advanced Materials. J.-L. Z. is supported by NSFC under Grant No.~12604193. Computing resources are provided by the SongShan Lake HPC Center (SSL-HPC) in Great Bay University.

\bibliography{apssamp}% Produces the bibliography via BibTeX.

@article{Basov2005RMP,
  title = {Electrodynamics of high-${T}_{c}$ superconductors},
  author = {Basov, D. N. and Timusk, T.},
  journal = {Rev. Mod. Phys.},
  volume = {77},
  issue = {2},
  pages = {721--779},
  numpages = {0},
  year = {2005},
  month = {Aug},
  publisher = {American Physical Society},
  doi = {10.1103/RevModPhys.77.721},
  url = {https://link.aps.org/doi/10.1103/RevModPhys.77.721}
}

@article{Kapitulnik2009NJP,
  title = {Polar Kerr effect as probe for time-reversal symmetry breaking in unconventional superconductors},
  author = {Kapitulnik, Aharon and Xia, Jing and Schemm, Elizabeth and Palevski, Alexander},
  journal = {New J. Phys.},
  volume = {11},
  issue = {5},
  pages = {055060},
  year = {2009},
  month = {May},
  publisher = {IOP Publishing},
  doi = {10.1088/1367-2630/11/5/055060},
  url = {https://iopscience.iop.org/article/10.1088/1367-2630/11/5/055060}
}

@article{Argyres1955PR,
  title = {Theory of the Faraday and Kerr Effects in Ferromagnetics},
  author = {Argyres, Petros N.},
  journal = {Phys. Rev.},
  volume = {97},
  issue = {2},
  pages = {334--345},
  numpages = {0},
  year = {1955},
  month = {Jan},
  publisher = {American Physical Society},
  doi = {10.1103/PhysRev.97.334},
  url = {https://link.aps.org/doi/10.1103/PhysRev.97.334}
}

@article{XiaJ2006PRL,
  title = {High Resolution Polar Kerr Effect Measurements of ${\mathrm{Sr}}_{2}{\mathrm{RuO}}_{4}$: Evidence for Broken Time-Reversal Symmetry in the Superconducting State},
  author = {Xia, Jing and Maeno, Yoshiteru and Beyersdorf, Peter T. and Fejer, M. M. and Kapitulnik, Aharon},
  journal = {Phys. Rev. Lett.},
  volume = {97},
  issue = {16},
  pages = {167002},
  numpages = {4},
  year = {2006},
  month = {Oct},
  publisher = {American Physical Society},
  doi = {10.1103/PhysRevLett.97.167002},
  url = {https://link.aps.org/doi/10.1103/PhysRevLett.97.167002}
}

@article{Schemm2014Science,
  title = {Observation of broken time-reversal symmetry in the heavy-fermion superconductor UPt$_3$},
  author = {Schemm, E. R. and Gannon, W. J. and Wishne, C. M. and Halperin, W. P. and Kapitulnik, Aharon},
  journal = {Science},
  volume = {345},
  issue = {6193},
  pages = {190--193},
  year = {2014},
  month = {Jul},
  publisher = {American Association for the Advancement of Science},
  doi = {10.1126/science.1248552},
  url = {https://www.science.org/doi/10.1126/science.1248552}
}

@article{Gong2017SciAdv,
  title = {Time-reversal symmetry-breaking superconductivity in epitaxial bismuth/nickel bilayers},
  author = {Gong, Xinxin and Kargarian, Mehdi and Stern, Alex and Yue, Di and Zhou, Hexin and Jin, Xiaofeng and Galitski, Victor M. and Yakovenko, Victor M. and Xia, Jing},
  journal = {Sci. Adv.},
  volume = {3},
  issue = {3},
  pages = {e1602579},
  year = {2017},
  month = {Mar},
  publisher = {American Association for the Advancement of Science},
  doi = {10.1126/sciadv.1602579},
  url = {https://www.science.org/doi/10.1126/sciadv.1602579}
}

@article{Levenson2018PRL,
  title = {Polar Kerr Effect from Time-Reversal Symmetry Breaking in the Heavy-Fermion Superconductor ${\mathrm{PrOs}}_{4}{\mathrm{Sb}}_{12}$},
  author = {Levenson-Falk, E. M. and Schemm, E. R. and Aoki, Y. and Maple, M. B. and Kapitulnik, A.},
  journal = {Phys. Rev. Lett.},
  volume = {120},
  issue = {18},
  pages = {187004},
  numpages = {6},
  year = {2018},
  month = {May},
  publisher = {American Physical Society},
  doi = {10.1103/PhysRevLett.120.187004},
  url = {https://link.aps.org/doi/10.1103/PhysRevLett.120.187004}
}

@article{Kallin2016RPP,
doi = {10.1088/0034-4885/79/5/054502},
url = {https://doi.org/10.1088/0034-4885/79/5/054502},
year = {2016},
month = {apr},
publisher = {IOP Publishing},
volume = {79},
number = {5},
pages = {054502},
author = {Kallin, Catherine and Berlinsky, John},
title = {Chiral superconductors},
journal = {Reports on Progress in Physics}
}

@article{Volovik1988jetp,
  title = {An analog of the quantum Hall effect in a superfluid 3He film},
  author = {Volovik, G.E.},
  journal = {Sov. Phys. JETP},
  volume = {67},
  issue = {9},
  pages = {1804},
  numpages = {0},
  year = {1988},
  month = {Apr},
  publisher = {},
  doi = {1988JETP...67.1804V},
  url = {https://www.jetp.ras.ru/cgi-bin/e/index/e/67/9/p1804?a=list}
}

@article{Read2000prb,
  title = {Paired states of fermions in two dimensions with breaking of parity and time-reversal symmetries and the fractional quantum Hall effect},
  author = {Read, N. and Green, Dmitry},
  journal = {Phys. Rev. B},
  volume = {61},
  issue = {15},
  pages = {10267--10297},
  numpages = {0},
  year = {2000},
  month = {Apr},
  publisher = {American Physical Society},
  doi = {10.1103/PhysRevB.61.10267},
  url = {https://link.aps.org/doi/10.1103/PhysRevB.61.10267}
}

@article{Yip1992JLTP,
title = {Circular dichroism and birefringence in unconventional superconductors},
author = {Yip, S. K. and Sauls, J. A.},
journal = {Journal of Low Temperature Physics},
volume = {86},
issue = {3},
pages = {257-290},
year = {1992},
url = {https://doi.org/10.1007/BF01151804},
doi = {10.1007/BF01151804}
}

@article{Roy2008PRB,
  title = {Collective modes and electromagnetic response of a chiral superconductor},
  author = {Roy, Rahul and Kallin, Catherine},
  journal = {Phys. Rev. B},
  volume = {77},
  issue = {17},
  pages = {174513},
  numpages = {7},
  year = {2008},
  month = {May},
  publisher = {American Physical Society},
  doi = {10.1103/PhysRevB.77.174513},
  url = {https://link.aps.org/doi/10.1103/PhysRevB.77.174513}
}

@article{Lutchyn2008PRB,
  title = {Gauge-invariant electromagnetic response of a chiral ${p}_{x}+i{p}_{y}$ superconductor},
  author = {Lutchyn, Roman M. and Nagornykh, Pavel and Yakovenko, Victor M.},
  journal = {Phys. Rev. B},
  volume = {77},
  issue = {14},
  pages = {144516},
  numpages = {21},
  year = {2008},
  month = {Apr},
  publisher = {American Physical Society},
  doi = {10.1103/PhysRevB.77.144516},
  url = {https://link.aps.org/doi/10.1103/PhysRevB.77.144516}
}

@article{Furusaki2001prb,
  title = {Spontaneous Hall effect in a chiral p-wave superconductor},
  author = {Furusaki, Akira and Matsumoto, Masashige and Sigrist, Manfred},
  journal = {Phys. Rev. B},
  volume = {64},
  issue = {5},
  pages = {054514},
  numpages = {17},
  year = {2001},
  month = {Jul},
  publisher = {American Physical Society},
  doi = {10.1103/PhysRevB.64.054514},
  url = {https://link.aps.org/doi/10.1103/PhysRevB.64.054514}
}

@article{Goryo2008prb,
  title = {Impurity-induced polar Kerr effect in a chiral $p$-wave superconductor},
  author = {Goryo, Jun},
  journal = {Phys. Rev. B},
  volume = {78},
  issue = {6},
  pages = {060501(R)},
  numpages = {4},
  year = {2008},
  month = {Aug},
  publisher = {American Physical Society},
  doi = {10.1103/PhysRevB.78.060501},
  url = {https://link.aps.org/doi/10.1103/PhysRevB.78.060501}
}

@article{Lutchyn2009prb,
  title = {Frequency and temperature dependence of the anomalous ac Hall conductivity in a chiral ${p}_{x}+i{p}_{y}$ superconductor with impurities},
  author = {Lutchyn, Roman M. and Nagornykh, Pavel and Yakovenko, Victor M.},
  journal = {Phys. Rev. B},
  volume = {80},
  issue = {10},
  pages = {104508},
  numpages = {21},
  year = {2009},
  month = {Sep},
  publisher = {American Physical Society},
  doi = {10.1103/PhysRevB.80.104508},
  url = {https://link.aps.org/doi/10.1103/PhysRevB.80.104508}
}

@article{Konig2017prl,
  title = {Kerr Effect from Diffractive Skew Scattering in Chiral ${p}_{x}\ifmmode\pm\else\textpm\fi{}i{p}_{y}$ Superconductors},
  author = {K\"onig, Elio J. and Levchenko, Alex},
  journal = {Phys. Rev. Lett.},
  volume = {118},
  issue = {2},
  pages = {027001},
  numpages = {5},
  year = {2017},
  month = {Jan},
  publisher = {American Physical Society},
  doi = {10.1103/PhysRevLett.118.027001},
  url = {https://link.aps.org/doi/10.1103/PhysRevLett.118.027001}
}

@article{LiY2020prr,
  title = {Anomalous Hall effect in single-band chiral superconductors from impurity superlattices},
  author = {Li, Yu and Wang, Zhiqiang and Huang, Wen},
  journal = {Phys. Rev. Res.},
  volume = {2},
  issue = {4},
  pages = {042027(R)},
  numpages = {6},
  year = {2020},
  month = {Nov},
  publisher = {American Physical Society},
  doi = {10.1103/PhysRevResearch.2.042027},
  url = {https://link.aps.org/doi/10.1103/PhysRevResearch.2.042027}
}

@article{LiuHT2023prb,
  title = {Impact of random impurities on the anomalous Hall effect in chiral superconductors},
  author = {Liu, Hao-Tian and Chen, Weipeng and Huang, Wen},
  journal = {Phys. Rev. B},
  volume = {107},
  issue = {22},
  pages = {224517},
  numpages = {9},
  year = {2023},
  month = {Jun},
  publisher = {American Physical Society},
  doi = {10.1103/PhysRevB.107.224517},
  url = {https://link.aps.org/doi/10.1103/PhysRevB.107.224517}
}

@article{Ahn2021NC,
  title = {Theory of optical responses in clean multi-band superconductors},
  author = {Ahn, Junyeong and Nagaosa, Naoto},
  journal = {Nature Communications},
  volume = {12},
  issue = {1},
  pages = {1617},
  year = {2021},
  doi = {10.1038/s41467-021-21905-x},
  url = {https://doi.org/10.1038/s41467-021-21905-x}
}

@article{GaoYang2026prb,
  title = {Pairing-induced momentum-space magnetism and its implication in optical anomalous Hall effect in chiral superconductors},
  author = {Geng, Bin and Gao, Yang and Niu, Qian},
  journal = {Phys. Rev. B},
  volume = {113},
  issue = {9},
  pages = {094527},
  numpages = {13},
  year = {2026},
  month = {Mar},
  publisher = {American Physical Society},
  doi = {10.1103/33b6-84kz},
  url = {https://link.aps.org/doi/10.1103/33b6-84kz}
}

@article{BrounD_prb2025,
  title = {Probing time-reversal symmetry breaking at microwave frequencies},
  author = {Chouinard, T. and Broun, D. M.},
  journal = {Phys. Rev. B},
  volume = {112},
  issue = {1},
  pages = {014521},
  numpages = {12},
  year = {2025},
  month = {Jul},
  publisher = {American Physical Society},
  doi = {10.1103/qwvx-xl1y},
  url = {https://link.aps.org/doi/10.1103/qwvx-xl1y}
}

@imsc{Broun_arXiv2025,
   title={Microwave resonator for measuring time-reversal symmetry breaking at cryogenic temperatures},
   volume={96},
   ISSN={1089-7623},
   url={http://dx.doi.org/10.1063/5.0293524},
   DOI={10.1063/5.0293524},
   number={11},
   journal={Review of Scientific Instruments},
   publisher={AIP Publishing},
   author={Chouinard, T. and Broun, D. M.},
   year={2025} 
}

@article{Niu1985PRB,
  title={Quantized Hall conductance as a topological invariant},
  author={Niu, Qian and Thouless, Ds J and Wu, Yong-Shi},
  journal={Physical Review B},
  volume={31},
  number={6},
  pages={3372},
  year={1985},
  publisher={APS},
doi = {10.1103/PhysRevB.31.3372}
}

@misc{WangZhi2026arXiv,
      title={Semiclassical theory for the orbital magnetic moment of superconducting quasiparticles}, 
      author={Jian-Hua Zeng and Zhongbo Yan and Zhi Wang and Qian Niu},
      year={2026},
      eprint={2602.17376},
      archivePrefix={arXiv},
      primaryClass={cond-mat.supr-con},
      url={https://arxiv.org/abs/2602.17376}, 
}

@misc{ZhuJ2026arXiv,
      title={Microscopic origin of orbital magnetization in chiral superconductors}, 
      author={Jihang Zhu and Chunli Huang},
      year={2026},
      eprint={2601.12387},
      archivePrefix={arXiv},
      primaryClass={cond-mat.supr-con},
      url={https://arxiv.org/abs/2601.12387v2}, 
}

@misc{Armitage2508,
      title={Terahertz range polarization rotation in the candidate time-reversal symmetry breaking superconductor BiNi}, 
      author={Ralph Romero III and Zhenisbek Tagay and Jiahao Liang and Jason Y. Yan and Di Yue and N. P. Armitage},
      year={2025},
      eprint={2508.08026},
      archivePrefix={arXiv},
      primaryClass={cond-mat.supr-con},
      url={https://arxiv.org/abs/2508.08026}, 
}

@misc{Yanase2026arXiv,
      title={Microwave Kerr/Faraday Resonance in Two-dimensional Chiral Superconductors}, 
      author={Taiki Matsushita and Jun'ichi Ieda and Yasufumi Araki and Takahiro Morimoto and Ilya Vekhter and Youichi Yanase},
      year={2026},
      eprint={2601.10151},
      archivePrefix={arXiv},
      primaryClass={cond-mat.supr-con},
      url={https://arxiv.org/abs/2601.10151}, 
}

@article{Huang2018PRL,
  title = {Possible Three-Dimensional Nematic Odd-Parity Superconductivity in ${\mathrm{Sr}}_{2}{\mathrm{RuO}}_{4}$},
  author = {Huang, Wen and Yao, Hong},
  journal = {Phys. Rev. Lett.},
  volume = {121},
  issue = {15},
  pages = {157002},
  numpages = {7},
  year = {2018},
  month = {Oct},
  publisher = {American Physical Society},
  doi = {10.1103/PhysRevLett.121.157002},
  url = {https://link.aps.org/doi/10.1103/PhysRevLett.121.157002}
}

@article{Huang2021PRR,
  title = {Possibility of mixed helical p-wave pairings in ${\mathrm{Sr}}_{2}{\mathrm{RuO}}_{4}$},
  author = {Huang, Wen and Wang, Zhiqiang},
  journal = {Phys. Rev. Res.},
  volume = {3},
  issue = {4},
  pages = {L042002},
  numpages = {7},
  year = {2021},
  month = {Oct},
  publisher = {American Physical Society},
  doi = {10.1103/PhysRevResearch.3.L042002},
  url = {https://link.aps.org/doi/10.1103/PhysRevResearch.3.L042002}
}

@article{ZhangJL2024PRL,
  title = {Quantum-Geometry-Induced Anomalous Hall Effect in Nonunitary Superconductors and Application to ${\mathrm{Sr}}_{2}{\mathrm{RuO}}_{4}$},
  author = {Zhang, Jia-Long and Chen, Weipeng and Liu, Hao-Tian and Li, Yu and Wang, Zhiqiang and Huang, Wen},
  journal = {Phys. Rev. Lett.},
  volume = {132},
  issue = {13},
  pages = {136001},
  numpages = {7},
  year = {2024},
  month = {Mar},
  publisher = {American Physical Society},
  doi = {10.1103/PhysRevLett.132.136001},
  url = {https://link.aps.org/doi/10.1103/PhysRevLett.132.136001}
}

@misc{Frilen2026arXiv,
      title={Quantum geometric anomalous Hall response in orbitally nonunitary superconductors}, 
      author={Viktor Frilen and Annica M. Black-Schaffer and Ankita Bhattacharya},
      year={2026},
      eprint={2606.13526},
      archivePrefix={arXiv},
      primaryClass={cond-mat.supr-con},
      url={https://arxiv.org/abs/2606.13526}, 
}

@article{RamiresJPCM2022,
doi = {10.1088/1361-648X/ac6d3a},
url = {https://doi.org/10.1088/1361-648X/ac6d3a},
year = {2022},
month = {jun},
publisher = {IOP Publishing},
volume = {34},
number = {30},
pages = {304001},
author = {Ramires, Aline},
title = {Nonunitary superconductivity in complex quantum materials},
journal = {Journal of Physics: Condensed Matter}
}

@article{FuL2015PRL,
  title = {Parity-Breaking Phases of Spin-Orbit-Coupled Metals with Gyrotropic, Ferroelectric, and Multipolar Orders},
  author = {Fu, Liang},
  journal = {Phys. Rev. Lett.},
  volume = {115},
  issue = {2},
  pages = {026401},
  numpages = {5},
  year = {2015},
  month = {Jul},
  publisher = {American Physical Society},
  doi = {10.1103/PhysRevLett.115.026401},
  url = {https://link.aps.org/doi/10.1103/PhysRevLett.115.026401}
}

@article{ChenWP2021PRR,
  title = {Quantum-geometry-induced intrinsic optical anomaly in multiorbital superconductors},
  author = {Chen, Weipeng and Huang, Wen},
  journal = {Phys. Rev. Res.},
  volume = {3},
  issue = {4},
  pages = {L042018},
  numpages = {5},
  year = {2021},
  month = {Nov},
  publisher = {American Physical Society},
  doi = {10.1103/PhysRevResearch.3.L042018},
  url = {https://link.aps.org/doi/10.1103/PhysRevResearch.3.L042018}
}

@article{Brydon2019PRX,
  title = {Loop Currents and Anomalous Hall Effect from Time-Reversal Symmetry-Breaking Superconductivity on the Honeycomb Lattice},
  author = {Brydon, P. M. R. and Abergel, D. S. L. and Agterberg, D. F. and Yakovenko, Victor M.},
  journal = {Phys. Rev. X},
  volume = {9},
  issue = {3},
  pages = {031025},
  numpages = {17},
  year = {2019},
  month = {Aug},
  publisher = {American Physical Society},
  doi = {10.1103/PhysRevX.9.031025},
  url = {https://link.aps.org/doi/10.1103/PhysRevX.9.031025}
}

@article{Can2021PRL,
  title = {Probing Time Reversal Symmetry Breaking Topological Superconductivity in Twisted Double Layer Copper Oxides with Polar Kerr Effect},
  author = {Can, Oguzhan and Zhang, Xiao-Xiao and Kallin, Catherine and Franz, Marcel},
  journal = {Phys. Rev. Lett.},
  volume = {127},
  issue = {15},
  pages = {157001},
  numpages = {6},
  year = {2021},
  month = {Oct},
  publisher = {American Physical Society},
  doi = {10.1103/PhysRevLett.127.157001},
  url = {https://link.aps.org/doi/10.1103/PhysRevLett.127.157001}
}

@article{HuYJ2025PRB,
  title = {Quantum geometric superfluid weight in multiband superconductors: A microscopic interpretation},
  author = {Hu, Yi-Jian and Huang, Wen},
  journal = {Phys. Rev. B},
  volume = {111},
  issue = {13},
  pages = {134511},
  numpages = {11},
  year = {2025},
  month = {Apr},
  publisher = {American Physical Society},
  doi = {10.1103/PhysRevB.111.134511},
  url = {https://link.aps.org/doi/10.1103/PhysRevB.111.134511}
}

@article{Taylor2012PRL,
  title = {Intrinsic Hall Effect in a Multiband Chiral Superconductor in the Absence of an External Magnetic Field},
  author = {Taylor, Edward and Kallin, Catherine},
  journal = {Phys. Rev. Lett.},
  volume = {108},
  issue = {15},
  pages = {157001},
  numpages = {5},
  year = {2012},
  month = {Apr},
  publisher = {American Physical Society},
  doi = {10.1103/PhysRevLett.108.157001},
  url = {https://link.aps.org/doi/10.1103/PhysRevLett.108.157001}
}

@article{Wysokinski2012PRL,
  title = {Intrinsic Optical Dichroism in the Chiral Superconducting State of ${\mathrm{Sr}}_{2}{\mathrm{RuO}}_{4}$},
  author = {Wysoki\ifmmode \acute{n}\else \'{n}\fi{}ski, K. I. and Annett, James F. and Gy\"orffy, B. L.},
  journal = {Phys. Rev. Lett.},
  volume = {108},
  issue = {7},
  pages = {077004},
  numpages = {5},
  year = {2012},
  month = {Feb},
  publisher = {American Physical Society},
  doi = {10.1103/PhysRevLett.108.077004},
  url = {https://link.aps.org/doi/10.1103/PhysRevLett.108.077004}
}

@article{Gradhand2013PRB,
  title = {Kerr rotation in the unconventional superconductor Sr${}_{2}$RuO${}_{4}$},
  author = {Gradhand, Martin and Wysokinski, Karol I. and Annett, James F. and Gy\"orffy, Balazs L.},
  journal = {Phys. Rev. B},
  volume = {88},
  issue = {9},
  pages = {094504},
  numpages = {8},
  year = {2013},
  month = {Sep},
  publisher = {American Physical Society},
  doi = {10.1103/PhysRevB.88.094504},
  url = {https://link.aps.org/doi/10.1103/PhysRevB.88.094504}
}

@article{Komendova2017PRL,
  title = {Odd-Frequency Superconductivity in ${\mathrm{Sr}}_{2}{\mathrm{RuO}}_{4}$ Measured by Kerr Rotation},
  author = {Komendov\'a, L. and Black-Schaffer, A. M.},
  journal = {Phys. Rev. Lett.},
  volume = {119},
  issue = {8},
  pages = {087001},
  numpages = {5},
  year = {2017},
  month = {Aug},
  publisher = {American Physical Society},
  doi = {10.1103/PhysRevLett.119.087001},
  url = {https://link.aps.org/doi/10.1103/PhysRevLett.119.087001}
}

@article{WangZQ2017PRB,
  title = {Intrinsic ac anomalous Hall effect of nonsymmorphic chiral superconductors with an application to ${\mathrm{UPt}}_{3}$},
  author = {Wang, Zhiqiang and Berlinsky, John and Zwicknagl, Gertrud and Kallin, Catherine},
  journal = {Phys. Rev. B},
  volume = {96},
  issue = {17},
  pages = {174511},
  numpages = {15},
  year = {2017},
  month = {Nov},
  publisher = {American Physical Society},
  doi = {10.1103/PhysRevB.96.174511},
  url = {https://link.aps.org/doi/10.1103/PhysRevB.96.174511}
}

@article{ZhangJL2020PRB,
  title = {Hidden anomalous Hall effect in ${\mathrm{Sr}}_{2}{\mathrm{RuO}}_{4}$ with chiral superconductivity dominated by the Ru ${d}_{xy}$ orbital},
  author = {Zhang, Jia-Long and Li, Yu and Huang, Wen and Zhang, Fu-Chun},
  journal = {Phys. Rev. B},
  volume = {102},
  issue = {18},
  pages = {180509(R)},
  numpages = {6},
  year = {2020},
  month = {Nov},
  publisher = {American Physical Society},
  doi = {10.1103/PhysRevB.102.180509},
  url = {https://link.aps.org/doi/10.1103/PhysRevB.102.180509}
}

@article{Denys2021PRB,
  title = {Origin of the anomalous Hall effect in two-band chiral superconductors},
  author = {Denys, M. D. E. and Brydon, P. M. R.},
  journal = {Phys. Rev. B},
  volume = {103},
  issue = {9},
  pages = {094503},
  numpages = {9},
  year = {2021},
  month = {Mar},
  publisher = {American Physical Society},
  doi = {10.1103/PhysRevB.103.094503},
  url = {https://link.aps.org/doi/10.1103/PhysRevB.103.094503}
}

@article{Taylor2013JPCM,
doi = {10.1088/1742-6596/449/1/012036},
url = {https://doi.org/10.1088/1742-6596/449/1/012036},
year = {2013},
month = {jul},
publisher = {},
volume = {449},
number = {1},
pages = {012036},
author = {Taylor, Edward and Kallin, Catherine},
title = {Anomalous Hall conductivity of clean Sr2RuO4 at finite temperatures},
journal = {Journal of Physics: Conference Series}
}

@article{Karplus1954PR,
  title = {Hall Effect in Ferromagnetics},
  author = {Karplus, Robert and Luttinger, J. M.},
  journal = {Phys. Rev.},
  volume = {95},
  issue = {5},
  pages = {1154--1160},
  numpages = {0},
  year = {1954},
  month = {Sep},
  publisher = {American Physical Society},
  doi = {10.1103/PhysRev.95.1154},
  url = {https://link.aps.org/doi/10.1103/PhysRev.95.1154}
}

@article{Provost1980,
    author = {Provost, J.P. and Vallee, G.},
    title = {Riemannian structure on manifolds of quantum states},
    journal = {Communications in Mathematical Physics},
    year = {1980},
    volume={76},
    issue={3},
    pages={289--301},
    doi={10.1007/BF02193559},
    url={https://doi.org/10.1007/BF02193559}
}

@article{Nagaosa2010RMP,
  title = {Anomalous Hall effect},
  author = {Nagaosa, Naoto and Sinova, Jairo and Onoda, Shigeki and MacDonald, A. H. and Ong, N. P.},
  journal = {Rev. Mod. Phys.},
  volume = {82},
  issue = {2},
  pages = {1539--1592},
  numpages = {0},
  year = {2010},
  month = {May},
  publisher = {American Physical Society},
  doi = {10.1103/RevModPhys.82.1539},
  url = {https://link.aps.org/doi/10.1103/RevModPhys.82.1539}
}

@article{XiaoD2010RMP,
  title = {Berry phase effects on electronic properties},
  author = {Xiao, Di and Chang, Ming-Che and Niu, Qian},
  journal = {Rev. Mod. Phys.},
  volume = {82},
  issue = {3},
  pages = {1959--2007},
  numpages = {0},
  year = {2010},
  month = {Jul},
  publisher = {American Physical Society},
  doi = {10.1103/RevModPhys.82.1959},
  url = {https://link.aps.org/doi/10.1103/RevModPhys.82.1959}
}

@article{Torma2023PRL,
  title={Essay: Where Can Quantum Geometry Lead Us?},
  author={T{\"o}rm{\"a}, P{\"a}ivi},
  journal={Physical Review Letters},
  volume={131},
  number={24},
  pages={240001},
  year={2023},
  publisher={APS},
  doi = {10.1103/PhysRevLett.131.240001},
  url = {https://link.aps.org/doi/10.1103/PhysRevLett.131.240001}
}

@article{LiuTY2024NSR,
  title={Quantum geometry in condensed matter},
  author={Liu, Tianyu and Qiang, Xiao-Bin and Lu, Hai-Zhou and Xie, X-C},
  journal={National Science Review},
  pages={nwae334},
  year={2024},
  publisher={Oxford University Press},
doi={10.1093/nsr/nwae334}
}

@article{YuJ2025NPJ,
    author = {Yu, Jiabin and Bernevig, B. Andrei and Queiroz, Raquel and Rossi, Enrico and Törmä, Päivi and Yang, Bohm-Jung},
    title = {Quantum geometry in quantum materials},
    journal = {npj Quantum Materials},
    year = {2025},
    volume={10},
    issue={1},
    pages={101},
    doi = {10.1038/s41535-025-00801-3},
    url = {https://doi.org/10.1038/s41535-025-00801-3}
}

@article{Cho2016PRL,
  title = {Necessity of Time-Reversal Symmetry Breaking for the Polar Kerr Effect in Linear Response},
  author = {Cho, Weejee and Kivelson, Steven A.},
  journal = {Phys. Rev. Lett.},
  volume = {116},
  issue = {9},
  pages = {093903},
  numpages = {5},
  year = {2016},
  month = {Mar},
  publisher = {American Physical Society},
  doi = {10.1103/PhysRevLett.116.093903},
  url = {https://link.aps.org/doi/10.1103/PhysRevLett.116.093903}
}

@article{HuLH2021PRB,
  title = {Spontaneous magnetization in unitary superconductors with time reversal symmetry breaking},
  author = {Hu, Lun-Hui and Wang, Xuepeng and Shang, T.},
  journal = {Phys. Rev. B},
  volume = {104},
  issue = {5},
  pages = {054520},
  numpages = {7},
  year = {2021},
  month = {Aug},
  publisher = {American Physical Society},
  doi = {10.1103/PhysRevB.104.054520},
  url = {https://link.aps.org/doi/10.1103/PhysRevB.104.054520}
}

@article{Mackenzie2017npjQM,
author  = {Mackenzie, Andrew P. and Scaffidi, Thomas and Hicks, Clifford W. and Maeno, Yoshiteru},
year  = {2017},
title = {Even odder after twenty-three years: the superconducting order parameter puzzle of Sr2RuO4},
journal  = {npj Quantum Materials},
pages  = {40},
volume  = {2},
URL  = {https://doi.org/10.1038/s41535-017-0045-4},
DOI  = {10.1038/s41535-017-0045-4}
}

@Article{Wysokinski2019CM,
AUTHOR = {Wysokiński, Karol Izydor},
TITLE = {Time Reversal Symmetry Breaking Superconductors: Sr2RuO4 and Beyond},
JOURNAL = {Condensed Matter},
VOLUME = {4},
YEAR = {2019},
NUMBER = {2},
ARTICLE-NUMBER = {47},
URL = {https://www.mdpi.com/2410-3896/4/2/47},
ISSN = {2410-3896},
DOI = {10.3390/condmat4020047}
}

@article{LiuY2021JSNM,
author  = {Leggett, Anthony J. and Liu, Ying},
year  ={2021},
title  ={ Symmetry Properties of Superconducting Order Parameter in Sr2RuO4},
journal  ={Journal of Superconductivity and Novel Magnetism},
pages  ={1647},
volume  ={34},
ISSUE  ={6},
URL  ={https://doi.org/10.1007/s10948-020-05717-6},
DOI  ={10.1007/s10948-020-05717-6}
}

@article{Huang2021CPB,
doi = {10.1088/1674-1056/ac2488},
url = {https://doi.org/10.1088/1674-1056/ac2488},
year = {2021},
month = {nov},
publisher = {Chinese Physical Society and IOP Publishing Ltd},
volume = {30},
number = {10},
pages = {107403},
author = {Huang, Wen},
title = {A review of some new perspectives on the theory of superconducting Sr2RuO4},
journal = {Chinese Physics B}
}

@article{Maeno2024JPSJ,
author = {Maeno ,Yoshiteru and Yonezawa ,Shingo and Ramires ,Aline},
title = {Still Mystery after All These Years —Unconventional Superconductivity of Sr2RuO4—},
journal = {Journal of the Physical Society of Japan},
volume = {93},
number = {6},
pages = {062001},
year = {2024},
doi = {10.7566/JPSJ.93.062001},
URL = { https://doi.org/10.7566/JPSJ.93.062001},
eprint = {https://doi.org/10.7566/JPSJ.93.062001}
}

@article{Maeno2024NP,
author  ={Maeno, Y. and Ikeda, A. and Mattoni, G.},
year  ={2024},
title  ={Thirty years of puzzling superconductivity in Sr2RuO4},
journal  ={Nature Physics},
pages  ={1712-1718},
volume  ={20},
ISSUE  ={11},
URI  ={https://doi.org/10.1038/s41567-024-02656-0},
DOI  ={10.1038/s41567-024-02656-0},
}

@article{GaoAnyuan2026RMP,
  title = {Quantum geometry phenomena in condensed matter systems},
  author = {Gao, Anyuan and Nagaosa, Naoto and Ni, Ni and Xu, Su-Yang},
  journal = {Rev. Mod. Phys.},
  volume = {98},
  issue = {3},
  pages = {035005},
  numpages = {67},
  year = {2026},
  month = {Sep},
  publisher = {American Physical Society},
  doi = {10.1103/f78t-ky69},
  url = {https://link.aps.org/doi/10.1103/f78t-ky69}
}

@article{ZengM2023PRB,
  title = {Spin-orbit coupled superconductivity with spin-singlet nonunitary pairing},
  author = {Zeng, Meng and Xu, Dong-Hui and Wang, Zi-Ming and Hu, Lun-Hui},
  journal = {Phys. Rev. B},
  volume = {107},
  issue = {9},
  pages = {094507},
  numpages = {16},
  year = {2023},
  month = {Mar},
  publisher = {American Physical Society},
  doi = {10.1103/PhysRevB.107.094507},
  url = {https://link.aps.org/doi/10.1103/PhysRevB.107.094507}
}

@misc{footnote1,
note = {In multiorbital models with finite spin-orbit coupling, spin polarization is generally accompanied by (atomic)-orbital polarization. The latter is sometimes associated with the so-called orbital-nonunitary pairing~\cite{Frilen2026arXiv}. In the present study, we will not distinguish between spin- and orbital-nonunitary pairings, as their intrinsic Hall responses can be described in a unified manner in our analysis.}
}

\clearpage
\appendix
\setcounter{figure}{0}
\setcounter{table}{0}
% 重定义编号格式，例如变为 A1, A2, ...
\renewcommand{\thefigure}{A\arabic{figure}}
\renewcommand{\thetable}{A\arabic{table}}
\begin{widetext}
\begin{center}
    \textbf{Appendices for: Theory of the intrinsic optical anomalous Hall responses in time-reversal symmetry breaking multiband superconductors} 

\text{Jia-Long Zhang}\textsuperscript{1} and \text{Wen Huang}\textsuperscript{1}\\[0.5ex]
\textsuperscript{1}\textit{School of Physical Sciences, Great Bay University, Dongguan 523000, China}
\end{center}

\section{Proof of vanishing Class A1 contribution in unitary pairings}
\label{app:A0}
In this section, we prove that the contribution of the Class A1 diagrams vanishes for unitary pairings. We first rewrite the general expression for the A1 diagram
\begin{equation}
\bV_{\bk}^{ab, s_1 s_2} \times \bV_{\bk}^{ba, s_3 s_4} g_{aa}^{s_1 s_4} \tilde{g}_{bb}^{s_2 s_3}
\label{eq:A1}
\end{equation}
There are $2^4=16$ distinct spin configurations for $\{s_p\}=\{s_{1}s_{2} s_{3}s_{4}\}$, which can be categorized into four types, as shown in Fig.~\ref{fig:A1-diagram}. The proof shall be carried out separately for different categories of the spin configuration, under the premise of unitary pairing. 

\begin{figure}[h]
\includegraphics[width=16cm]{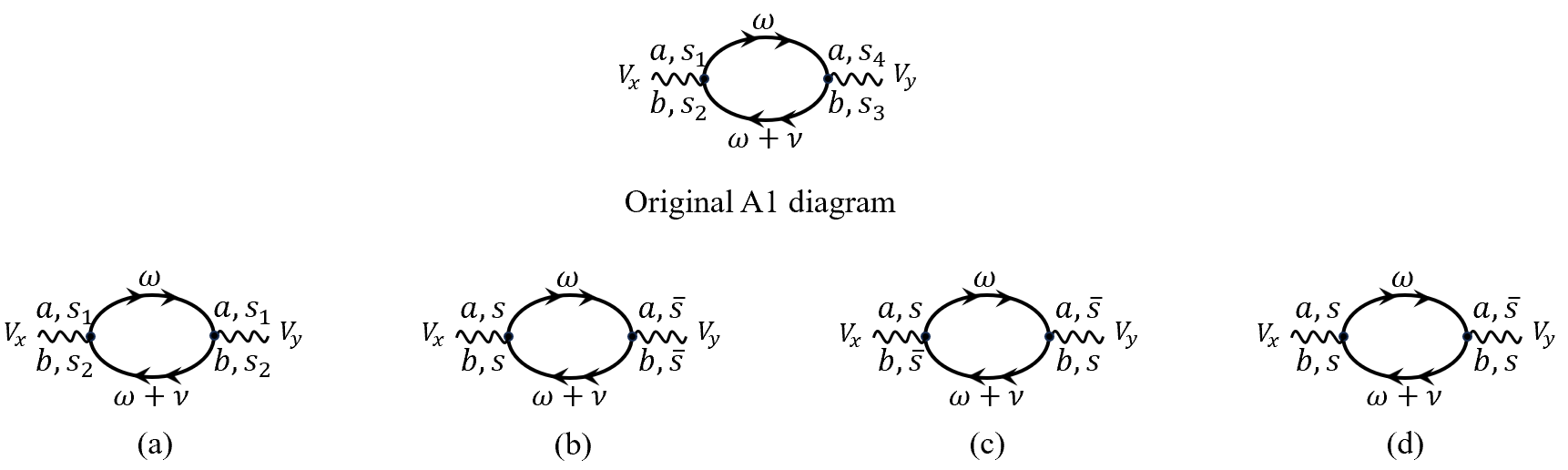}
\caption{The original Class A1 diagram (upper panel) and its various forms exhibiting different configurations of spin indices $\{s_p\}$ (lower panel).}
\label{fig:A1-diagram}
\end{figure}  

\begin{enumerate}
\item  $s_1 = s_4, \quad s_2 = s_3$. 

This scenario corresponds to Fig.~\ref{fig:A1-diagram}~(a), which encupsulates four spin configurations: $\{\ua\ua\ua\ua \}, \{\ua\da\da\ua \},  \{\da\ua\ua\da \}, \\ \{\da\da\da\da \}$. In this case, Eq.~\eqref{eq:A1} becomes
\begin{equation}
     \bV_{\bk}^{ab, s_1 s_2} \times \bV_{\bk}^{ba, s_2 s_1} g_{aa}^{s_1 s_1} \tilde{g}_{bb}^{s_2 s_2}  \,,
\end{equation}
by TRS, inversion symmetry and hermiticity, we have $\bV_{\bk}^{ab, s_1 s_2} = \bV_{\bk}^{ba, \bar{s}_2 \bar{s}_1}$ (see Eq.~(13)), thus $\bV_{\bk}^{ab, s_1 s_2} \times \bV_{\bk}^{ba, s_2 s_1} 
    = - \bV_{\bk}^{ab, \bar{s}_1 \bar{s}_2} \times \bV_{\bk}^{ba, \bar{s}_2 \bar{s}_1}$.
On the other hand, for unitary pairing states one has $g_{aa}^{s_1 s_1} \tilde{g}_{bb}^{s_2 s_2} = g_{aa}^{\bar{s}_1 \bar{s}_1} \tilde{g}_{bb}^{\bar{s}_2 \bar{s}_2}$.
It follows that this term vanishes for unitary states after summing over the spin indices.

\item $s_1 = \bar{s}_4, \quad s_2 = \bar{s}_3, \quad s_1 = s_2 = s$.

The corresponding diagram is shown in Fig.~\ref{fig:A1-diagram}~(b), which contains two spin configurations: $\{\ua\ua\da\da \}, \{\da\da\ua\ua \}$. Equation~\eqref{eq:A1} now reads
\begin{equation*}
    \bV_{\bk}^{ab, s s} \times \bV_{\bk}^{ba, \bar{s} \bar{s}} g_{aa}^{s \bar{s}} \tilde{g}_{bb}^{s \bar{s}} \,,
\end{equation*}
    
Using Eq.~(13), we obtain $\bV_{\bk}^{ab, s s} \times \bV_{\bk}^{ba, \bar{s} \bar{s}} = - \bV_{\bk}^{ab, ss} \times \bV_{\bk}^{ba, \bar{s} \bar{s}}$, implying that this contribution also vanishes.

\item  $s_1 = s_3 = s, \quad s_2 = s_4 = \bar{s}$.

This condition corresponds to Fig.~\ref{fig:A1-diagram} (c), which contains the following two spin configurations:   $\{\ua\da\ua\da \},, \{\da\ua\da\ua \}$. In this case, Eq.~\eqref{eq:A1} is given by
    \begin{equation*}
     \bV_{\bk}^{ab, s \bar{s}} \times \bV_{\bk}^{ba, s \bar{s}} g_{aa}^{s \bar{s}} \tilde{g}_{bb}^{ \bar{s}s} \,,
\end{equation*}
    
Utilizing Eq.~(13), one obtains $\bV_{\bk}^{ab, s \bar{s}} \times \bV_{\bk}^{ba, s \bar{s}} = \bV_{\bk}^{ba, s \bar{s}} \times \bV_{\bk}^{ab, s \bar{s}}  =- \bV_{\bk}^{ab, s \bar{s}} \times \bV_{\bk}^{ba, s \bar{s}} = 0 $. Hence, the contribution from these diagrams also vanishes.

\item  One out of four spin indices is opposite to the remaining three. 

This condition is represented by Fig.~\ref{fig:A1-diagram}~(d), and there are in total eight different spin configurations:   $\{\ua\ua\ua\da\}, \{\ua\ua\da\ua\}, \{\ua\da\ua\ua\}, \{\da\ua\ua\ua\}, \{\da\da\da\ua\}, \{\da\da\ua\da\}, \{\da\ua\da\da\}, \{\ua\da\da\da\}$. In this case, we consider the summation of the following two terms,
\begin{align}
     \bV_{\bk}^{ab, s s} \times \bV_{\bk}^{ba, s \bar{s}} g_{aa}^{s \bar{s}} \tilde{g}_{bb}^{s s} + \bV_{\bk}^{ab, s\bar{s} } \times \bV_{\bk}^{ba, \bar{s} \bar{s}} g_{aa}^{s \bar{s}} \tilde{g}_{bb}^{\bar{s}\bar{s}}  
    = \bV_{\bk}^{ab, s s} \times \bV_{\bk}^{ba, s \bar{s}} (g_{aa}^{s \bar{s}} \tilde{g}_{bb}^{s s} -g_{aa}^{s \bar{s}} \tilde{g}_{bb}^{\bar{s}\bar{s}} )  =0 \,,
\end{align}
where we have used $   \bV_{\bk}^{ab, s\bar{s} } \times \bV_{\bk}^{ba, \bar{s} \bar{s}}  = -\bV_{\bk}^{ab, s s} \times \bV_{\bk}^{ba, s \bar{s}}$ and for unitary state $ \tilde{g}_{bb}^{s s}=\tilde{g}_{bb}^{\bar{s} \bar{s}}$.
\end{enumerate}
Note that the above analyses hold irrespective of whether the pairing belongs to spin-triplet or spin-singlet, nor whether interband pairing is present. Hence, we have proven that the contribution of the A1 diagram to the anomalous Hall response vanishes identically for any unitary pairing states.

\section{Class A1 contribution in nonunitary pairings}
\label{app:A}
In this section, we present a model that illustrates the Class A1 contribution to the anomalous Hall conductivity in the nonunitary state. It is a minimal two-band model, which has vanishing interband pairing and where Cooper pairing takes place only between equal pseudospins in the band-basis representation. The BdG Hamiltonian matrix for the spin-$s$ component is given by
\begin{equation}
\label{eq:Ham_intra}
H_{\bk s}^{\text{sc}} = \begin{pmatrix}
\epsilon^{s}_{a\bk} & & \Delta_{aa, \bk}^{ss} & \\
 & \epsilon^{s}_{b\bk} & & \Delta_{bb, \bk}^{ss} \\
(\Delta_{aa,\bar{\bk}}^{ss})^{*} & & -\epsilon^{s}_{a\bar{\bk}} & \\
 & (\Delta_{bb,\bk}^{ss})^{*} & & -\epsilon^{s}_{b\bar{\bk}}
\end{pmatrix} \,,
\end{equation}
where $\epsilon_{a(b) \bk}$ is the normal state band dispersion, and $\Delta_{aa,\bk}^{ss}$, $\Delta_{bb,\bk}^{ss}$ are the intraband pairing gap functions on band-$a$ and band-$b$, respectively. The corresponding Hall conductivity can be obtained directly via the spectral representation and reads as follows
\begin{eqnarray}
\label{eq:ahcs}
\sigma_{H} (\omega) &=& -\frac{i}{2}\sum_{s,s^\prime}\sum_{\bk} \frac{ \epsilon^{s}_{a\bk }/E_{a\bk}^{s}-\epsilon^{s^{\prime}}_{b\bk }/E_{b\bk }^{s^\prime}}{(E_{a\bk}^{s}+E_{b\bk}^{s^\prime})^2-(\omega+i\delta)^2}  \left[ \bV^{ab,ss^\prime}_{\bk}\times \bV^{ba,s^\prime s}_{\bk}+(\bV^{ba,ss^\prime}_{\bar{\bk}})^{*}\times (\bV^{ab,s^\prime s}_{\bar{\bk}})^{*} \right]  \nonumber \\
 &=&  -i\sum_{s,s^\prime}\sum_{\bk} \frac{ \left(\epsilon^{s}_{a\bk }/E_{a\bk}^{s}-\epsilon^{s^{\prime}}_{b\bk }/E_{b\bk }^{s^\prime} \right) \bV^{ab,ss^\prime}_{\bk}\times \bV^{ba,s^\prime s}_{\bk}}{(E_{a\bk}^{s}+E_{b\bk}^{s^\prime})^2-(\omega+i\delta)^2} \,,
\end{eqnarray}
where $E^{s}_{i\bk}= \sqrt{(\epsilon^{s}_{i\bk})^2 + |\Delta_{ii,\bk}^{ss}|^2},~i=a,b$. In obtaining the second line, we have used the hermiticity and inversion symmetry conditions that render $(\bV^{ba,ss^\prime}_{\bar{\bk}})^{*}\times (\bV^{ab,s^\prime s}_{\bar{\bk}})^{*} = \bV^{ab,ss^\prime}_{\bk}\times \bV^{ba,s^\prime s}_{\bk}$. The form of the velocity cross product in the first line suggests that only Class A1 contribution is present. 

For nonunitary states, $E^s_{i\bk} \neq E^{\bar{s}}_{i\bk}$, it is readily seen that the Hall conductivity is generally finite. In the special case where $V^{ab,s\bar{s}}_{\bk} \equiv 0$, Eq.~\eqref{eq:ahcs1} can be further simplified to 
\begin{equation}
\label{eq:ahcs1}
\sigma^{s}_{H} (\omega) = -i\sum_{\bk} \frac{ \left(\epsilon^{s}_{a\bk }/E_{a\bk}^{s}-\epsilon^{s}_{b\bk }/E_{b\bk }^{s}\right) \bV^{ab,ss}_{\bk}\times \bV^{ba,ss}_{\bk} }{(E_{a\bk}^{s}+E_{b\bk}^{s})^2-(\omega+i\delta)^2} \,.
\end{equation}
The expression $\bV^{ab,ss}_{\bk}\times \bV^{ba,ss}_{\bk}$ describes a quantity proportional to the Berry curvature of the pseudospin-$s$ Bloch state (see Eq.~(6) in maintext). Owing to the normal-state time-reversal symmetry, this Berry curvature vanishes in the absence of SOC. Hence, our result also highlights the critical role of SOC in causing the intrinsic Hall effect in nonunitary states. Time-reversal and inversion symmetries further imply
\begin{equation}
\bV^{ab,\bar{s}\bar{s}}_{\bk}\times \bV^{ba,\bar{s}\bar{s}}_{\bk} = -\bV^{ab,ss}_{\bk}\times \bV^{ba,ss}_{\bk} \,.
\end{equation}
Furthermore, using the electron occupation number of the individual spin species $\rho^s_{i\bk} = \frac{1}{2}(1-\epsilon^s_{i\bk}/E^s_{i\bk})$, $\epsilon^{s}_{a\bk }/E_{a\bk}^{s}-\epsilon^{s}_{b\bk }/E_{b\bk }^{s}=2(\rho^s_{b\bk}-\rho^s_{a\bk})$. Put together, at large frequency the Hall conductivity Eq.~\eqref{eq:ahcs1} can be approximated by
\begin{equation}
\label{eq:ahcs2}
\sigma_{H} (\omega) \sim \frac{2i}{\omega^2}\sum_{\bk} [(\rho^\ua_{b\bk} - \rho^\da_{b\bk}) -(\rho^\ua_{a\bk} - \rho^\da_{a\bk}) ]\left[ \bV^{ab,\ua\ua}_{\bk}\times \bV^{ba,\ua\ua}_{\bk}+(\bV^{ba,\ua\ua}_{\bar{\bk}})^{*}\times (\bV^{ab,\ua\ua}_{\bar{\bk}})^{*} \right] \,.
\end{equation}
It can be seen that the high-frequency anomalous Hall conductivity is linearly related to the difference of the spin polarization in the two bands.

\section{Class A2 and B1 contributions}
\label{app:B}
In this appendix, we derive the anomalous Hall conductivity and classify its contributing terms for a spinless two-band model with unitary TRSB pairing. The goal is to confirm that the Hall response of such a model is associated with the Class A2 and B1 diagrams. This derivation can be readily extended to spinful unitary pairing states. Moreover, the analysis can also be applied to the subset of nonunitary states whose spin-up and spin-down sectors fully decouple in the Hamiltonian.  

The band-basis Hamiltonian is given by $H=\sum_{\bk} \Psi^\dagger_{\bk} \hat{H}_{\bk}^{\text{sc}}\Psi_{\bk}$ with $\Psi_{\bk}^T=(c_{a\bk},c_{b\bk},c^{\dagger}_{a\bar{\bk}},c^{\dagger}_{b\bar{\bk}})$ and 
\begin{equation}
H_{\bk}^{\text{sc}} = \begin{pmatrix}
\epsilon_{a\bk} & & \Delta_{aa,\bk} & \Delta_{ab,\bk}  \\
& \epsilon_{b\bk}  & \Delta_{ab,\bk} & \Delta_{bb,\bk} \\
\Delta^{*}_{aa,\bk} & \Delta^{*}_{ab,\bk} & -\epsilon_{a\bk} & \\
\Delta^{*}_{ab,\bk} & \Delta^{*}_{bb,\bk} & &  -\epsilon_{b\bk}
\end{pmatrix}  \,,
\end{equation}
The corresponding Green's function is expressed as
\begin{equation}
    G(\bk,i\omega_n) =(i\omega_{n}-H_{\bk}^{\text{sc}})^{-1}=\begin{pmatrix}
        g_{aa}  & g_{ab} & f_{aa} & f_{ab} \\
        g_{ba}  & g_{bb} & f_{ba} & f_{bb} \\
        f^{*}_{aa}  & f^{*}_{ab} & -g^{*}_{aa} & -g^{*}_{ab} \\
          f^{*}_{ba}  & f^{*}_{bb} & -g^{*}_{ba} & -g^{*}_{bb}
        \end{pmatrix}  \,.
\end{equation}
Note that, for simplicity, we have suppressed the explicit momentum and frequency dependence for the Green's function matrix elements. These elements are given explicitly by
\begin{align}
    g_{aa}(\bk,i\omega_{n}) &= \frac{(i \omega_{n}+\epsilon_{a})\left[(i\omega_{n})^2-(\epsilon_{b}^2+\Delta_{bb}^2)\right]-|\Delta_{ab}|^{2}(i\omega_{n}+\epsilon_{b})}{\text{Det}(G^{-1})} \,, \\
    g_{bb}(\bk,i\omega_{n}) &= \frac{[(i \omega_{n}+\epsilon_{b})\left[(i\omega_{n})^2-(\epsilon_{a}^2+\Delta_{aa}^2)\right]-|\Delta_{ab}|^{2}(i\omega_{n}+\epsilon_{a})}{\text{Det}(G^{-1})} \,, \\
    g_{ab}(\bk,i\omega_{n}) &= \frac{(i\omega_{n}+\epsilon_{a})\Delta_{bb}^{*}\Delta_{ab}+(i\omega_{n}+\epsilon_{b})\Delta_{ab}^{*}\Delta_{aa}}{\text{Det}(G^{-1})} \,, \\
    g_{ba}(\bk,i\omega_{n}) &= \frac{(i\omega_{n}+\epsilon_{b})\Delta_{aa}^{*}\Delta_{ab}+(i\omega_{n}+\epsilon_{a})\Delta_{ab}^{*}\Delta_{bb}}{\text{Det}(G^{-1})}  \,, \\
    f_{aa}(\bk,i\omega_{n}) &= \frac{[(i\omega_{n})^2-\epsilon_{b}^2]\Delta_{aa}+\Delta_{bb}^{*}[(\Delta_{ab})^2-\Delta_{aa}\Delta_{bb}]}{\text{Det}(G^{-1})} \,, \\
    f_{bb}(\bk,i\omega_{n}) &= \frac{[(i\omega_{n})^2-\epsilon_{a}^2]\Delta_{bb}+\Delta_{aa}^{*}[(\Delta_{ab})^2-\Delta_{aa}\Delta_{bb}]}{\text{Det}(G^{-1})} \,, \\
    f_{ab}(\bk,i\omega_{n}) &= \frac{(i\omega_{n}+\epsilon_{a})(i\omega_{n}-\epsilon_{b})\Delta_{ab}-\Delta_{ab}^{*}[(\Delta_{ab})^2-\Delta_{aa}\Delta_{bb}]}{\text{Det}(G^{-1})} \,, \\
    f_{ba}(\bk,i\omega_{n}) &= \frac{(i\omega_{n}-\epsilon_{a})(i\omega_{n}+\epsilon_{b})\Delta_{ab}-\Delta_{ab}^{*}[(\Delta_{ab})^2-\Delta_{aa}\Delta_{bb}]}{\text{Det}(G^{-1})}  \,.
\end{align}
Here $\text{Det}(G^{-1})= [(i\omega_{n})^2-E_{a}^2)][(i\omega_{n})^2-E_{b}^2]$, and $E_{a}$, $E_{b}$ are the two positive branches of the Bogoliubov dispersion. Explicit calculation of the Hall conductivity Eq.~(1) in the maintext then results in a series of terms, as classified in Table.~\ref{tab:A2-B1} according to the type of microscopic processes from which they originate. It can be shown that only the terms associated with Class A2 and B1 processes can be finite. Figure \ref{fig:A2-B1-diagram} shows some of the representative contributing diagrams. The complete form of the Hall conductivity acquires the following compact form
\begin{equation}
    \sigma_{H}(i\nu_{m})=\frac{2\nu_{m}T}{\Omega}\sum_{\bk,\omega_{n}}\frac{(\nu_{m}+2\omega_{n})^2}{\text{Det}(G^{-1})\text{Det}(\tilde{G}^{-1})}\left[ \epsilon_{a}\text{Im}(\Delta_{bb}^{*}\Delta_{ab})- \epsilon_{b}\text{Im}(\Delta_{aa}^{*}\Delta_{ab}) \right] (\bV_{\bk}^{aa}-\bV_{\bk}^{bb})\times\bV_{\bk}^{ab} \,.
\end{equation}
Notice that this result agrees with that obtained in Ref.~\onlinecite{Taylor2012PRL}, had the latter been performed in the band-basis representation as we did here. 

Extending the above derivations to spinful unitary TRSB pairings and nonunitary states, we readily see that SOC is not a prerequisite for this mechanism to manifest, unlike for the Class A1 diagrams.

\begin{table}[h]
    \centering
    \begin{tabular}{c|c}
    \hline
$\begin{aligned}
&(g_{aa}\tilde{g}_{ba}-g_{aa}\tilde{g}_{aa})\bV_{\bk}^{aa} \times\bV_{\bk}^{bb}
\end{aligned}$ &  ($\times$)  \\ \hline
$\begin{aligned}
&(g_{bb}\tilde{g}_{aa}-g_{aa}\tilde{g}_{bb})\bV_{\bk}^{ab} \times \bV_{\bk}^{ba}  
\end{aligned}$ &  Class A1 ($\times$)\\ \hline

$\begin{aligned}
&(g_{aa}\tilde{g}_{ba}-g_{ba}\tilde{g}_{aa})\bV_{\bk}^{aa}\times\bV_{\bk}^{ab} \\
&(g_{ab}\tilde{g}_{aa}-g_{aa}\tilde{g}_{ab})\bV_{\bk}^{aa}\times\bV_{\bk}^{ba}  \\ 
&(g_{ba}\tilde{g}_{bb}-g_{bb}\tilde{g}_{ba})\bV_{\bk}^{bb}\times\bV_{\bk}^{ab} \\ 
&(g_{bb}\tilde{g}_{ab}-g_{ab}\tilde{g}_{bb})\bV_{\bk}^{bb}\times\bV_{\bk}^{ba}
\\ \hline 
&\left[(g_{aa}\tilde{g}_{ba}-g_{ba}\tilde{g}_{aa})\bV_{\bar{\bk}}^{aa}\times\bV_{\bar{\bk}}^{ab}\right]^{*} 
\\ 
&\left[(g_{ab}\tilde{g}_{aa}-g_{aa}\tilde{g}_{ab})\bV_{\bar{\bk}}^{aa}\times\bV_{\bar{\bk}}^{ba}\right]^{*}   \\ 
&\left[(g_{ba}\tilde{g}_{bb}-g_{bb}\tilde{g}_{ba})\bV_{\bar{\bk}}^{bb}\times\bV_{\bar{\bk}}^{ab}\right]^{*}  \\ 
&\left[(g_{bb}\tilde{g}_{ab}-g_{ab}\tilde{g}_{bb})\bV_{\bar{\bk}}^{bb}\times\bV_{\bar{\bk}}^{ba}\right]^{*} 
\end{aligned}$
& Class A2  ($\checkmark$)\\  \hline
$\begin{aligned} 
&-(f_{aa}\tilde{f}^{*}_{aa}-f^{*}_{aa}\tilde{f}_{aa})\bV_{\bk}^{aa}\times(\bV_{\bar{\bk}}^{aa})^{*} \\ 
&-(f_{ab}\tilde{f}^{*}_{ba}-f^{*}_{ba}\tilde{f}_{ab})\bV_{\bk}^{aa}\times(\bV_{\bar{\bk}}^{bb})^{*} \\
&-(f_{ba}\tilde{f}^{*}_{ab}-f^{*}_{ab}\tilde{f}_{ba})\bV_{\bk}^{bb}\times(\bV_{\bar{\bk}}^{aa})^{*} \\ 
&-(f_{bb}\tilde{f}^{*}_{bb}-f^{*}_{bb}\tilde{f}_{bb})\bV_{\bk}^{bb}\times(\bV_{\bar{\bk}}^{bb})^{*} 
\end{aligned}$
& ($\times$)\\  \hline

$\begin{aligned} 
&-(f_{ab}\tilde{f}^{*}_{aa}-f^{*}_{aa}\,\tilde{f}_{ab})\bV_{\bk}^{aa}\times(\bV_{\bar{\bk}}^{ab})^{*} \\ 
&-(f_{aa}\,\tilde{f}^{*}_{ba}-f^{*}_{ba}\,\tilde{f}_{aa})\bV_{\bk}^{aa}\times(\bV_{\bar{\bk}}^{ba})^{*} \\ 
&-(f_{bb}\,\tilde{f}^{*}_{ab}-f^{*}_{ab}\,\tilde{f}_{bb})\bV_{\bk}^{bb}\times(\bV_{\bar{\bk}}^{ab})^{*} \\ 
&-(f_{ba}\tilde{f}^{*}_{bb}-f^{*}_{bb}\tilde{f}_{ba})\bV_{\bk}^{bb}\times(\bV_{\bar{\bk}}^{ba})^{*} \\ \hline 
&-(f_{ba}\tilde{f}^{*}_{aa}-f^{*}_{aa}\tilde{f}_{ba} )\bV_{\bk}^{ab}\times(\bV_{\bar{\bk}}^{aa})^{*} \\ 
&-(f_{bb}\tilde{f}^{*}_{ba}-f^{*}_{ba}\tilde{f}_{bb})\bV_{\bk}^{ab}\times(\bV_{\bar{\bk}}^{bb})^{*} \\ 
&-(f_{aa}\tilde{f}^{*}_{ab}-f^{*}_{ab}\tilde{f}_{aa})\bV_{\bk}^{ba}\times(\bV_{\bar{\bk}}^{aa})^{*} \\ 
&-(f_{ab}\tilde{f}^{*}_{bb}-f^{*}_{bb}\tilde{f}_{ab})\bV_{\bk}^{ba}\times(\bV_{\bar{\bk}}^{bb})^{*} \end{aligned}$
& Class B1 ($\checkmark$)\\  \hline

$\begin{aligned} &-(f_{bb}\,\tilde{f}^{*}_{aa}-f^{*}_{aa}\tilde{f}_{bb})\, \bV_{\bk}^{ab}\times(\bV_{\bar{\bk}}^{ab})^{*} \\
&-(f_{aa}\tilde{f}^{*}_{bb}-\tilde{f}_{aa}f^{*}_{bb})\bV_{\bk}^{ba}\times(\bV_{\bar{\bk}}^{ba})^{*} \end{aligned}$
& Class B2 ($\times$)\\  \hline

$\begin{aligned}  &-(f_{ba}\tilde{f}^{*}_{ba}-f^{*}_{ba}\tilde{f}_{ba})\bV_{\bk}^{ab}\times(\bV_{\bar{\bk}}^{ba})^{*} \\ 
&-(f_{ab}\tilde{f}^{*}_{ab}-f^{*}_{ab}\tilde{f}_{ab})\bV_{\bk}^{ba}\times(\bV_{\bar{\bk}}^{ab})^{*} \end{aligned}$
& Class B3 ($\times$)\\  \hline 
    \end{tabular}
    \caption{Classification of the terms that appear in the Hall conductivity of the spinless two-band model with unitary pairing in App.~\ref{app:B}. The `$\times$' symbol indicates that the corresponding terms vanish, while the `$\checkmark$' symbol marks contributions that can generally emerge.}
    \label{tab:A2-B1}
\end{table}

\begin{figure}
\includegraphics[width=16cm]{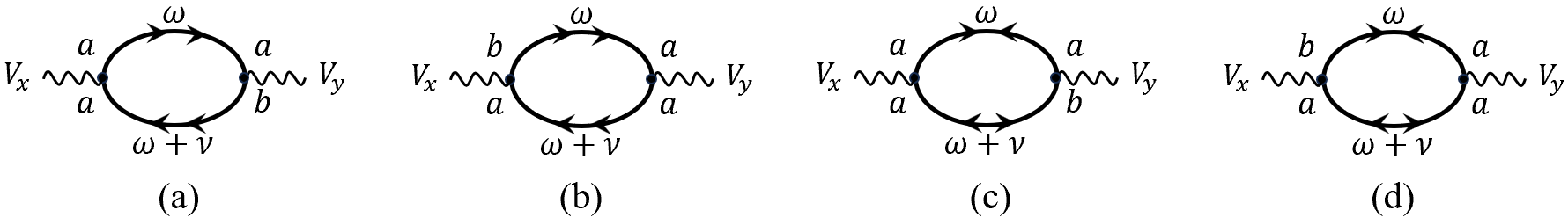}
\caption{(a) and (b): two typical Class A2 Feynman diagrams; (c) and (d): two typical Class B1 Feynman diagrams for unitary superconducting states. Note the difference in the position of the band index ``$b$". Again, the counter diagrams with the velocity vertices interchanged are not shown.}
\label{fig:A2-B1-diagram}
\end{figure}

\section{Effective electronic theory for nonunitary pairing}
\label{app:EETnonunitary}
In this section, we derive the effective electronic theory for nonunitary superconducting states. To illustrate the key physics, we consider a minimal two-band nonunitary model with the two pseudospin species fully decoupled in the Hamiltonian, i.e., $\Delta^{s\bar{s}}=0$ and $\bV_{\bk}^{s\bar{s}}=0$ and $\bV^{ab,s\bar{s}}=0$. Consider the following general pairing matrix for spin-$s$,
\begin{equation}
\hat{\Delta}_{\bk s} = \begin{pmatrix}
\Delta^{ss}_{aa,\bk}  &  \Delta^{ss}_{ab,\bk} \\ 
\Delta^{ss}_{ba,\bk} &  \Delta^{ss}_{bb,\bk} 
\end{pmatrix} \,.
\end{equation}
Projecting out the hole sector following the same procedure as in the derivation for unitary pairing returns an effective electron-only Hamiltonian that is also block diagonal in spin,
\begin{equation}
\hat{H}^{\prime}_{\bk s}=\begin{pmatrix}
\epsilon_{a\bk s}^\prime  &  \lambda_{\bk s}  \\
\lambda_{\bk s}^\ast &  \epsilon_{b\bk s}^\prime
\end{pmatrix} 
\label{eq:effHelectron1}
\end{equation}
with the spin-dependent renormalized dispersion and effective interband hybridization given by
\begin{eqnarray}
\lambda_{\bk s} &=& \frac{(\Delta_{aa,\bk}^{ss})^\ast\Delta^{ss}_{ab,\bk}}{2\epsilon_{a\bk}}+\frac{(\Delta^{ss}_{ba,\bk})^\ast\Delta^{ss}_{bb,\bk}}{2\epsilon_{b\bk}} \,, \nonumber \\
\epsilon_{a\bk s}^\prime &= & \epsilon_{a\bk}+\frac{|\Delta^{ss}_{aa,\bk}|^2}{2\epsilon_{a\bk}}+\frac{|\Delta^{ss}_{ab,\bk}|^2}{2\epsilon_{b\bk}} \,, \nonumber \\
\epsilon^\prime_{b\bk s}&=&\epsilon_{b\bk}+\frac{|\Delta^{ss}_{bb,\bk}|^2}{2\epsilon_{b\bk}}+\frac{|\Delta^{ss}_{ab,\bk}|^2}{2\epsilon_{a\bk}} \,.
\end{eqnarray}
Using the original normal-state velocity operators and the corrected electronic Green's function $(i\omega_n - \hat{H}^\prime_{\bk s})^{-1}$, the Hall conductivity of the spin-$s$ part of the effective model can be shown to be given by,
\begin{equation}
\sigma_{H }^{s}(i\nu_{m})  =\sum_{\bk,\omega_{n},s}\frac{i \text{Im}[\lambda_{\bk s}(\bV_{\bk}^{aa,ss}-\bV_{\bk}^{bb,ss})\times \bV_{\bk}^{ba,ss}] +(\epsilon^{\prime}_{a\bk s}-\epsilon^{\prime}_{b\bk s})\bV^{ab, ss}_{\bk}\times \bV_{\bk}^{ba,ss} }{2[(i\omega_n-\frac{\epsilon^\prime_{a\bk s}+\epsilon^\prime_{b\bk s}}{2})^2-E_{\bk s}^2 ] [(i\omega_n+i\nu_m-\frac{\epsilon^\prime_{a\bk s}+\epsilon^\prime_{b\bk s}}{2})^2-E_{\bk s}^2 ]} \,,
\label{eq:AHCeffectiveTheory2}  
\end{equation}
where $E_{\bk s}=\sqrt{(\frac{\epsilon_{a\bk s}^{\prime}-\epsilon_{b\bk s}^{\prime}}{2})^2+|\lambda_{\bk s}|^2}$. 

In the above expression, the first term is similar to that for unitary states, i.e., Eq.~(28) in the maintext. Hence, it captures the same microscopic contributions to the Hall response as in the latter, requiring both intraband and interband pairings. The second term is unique to nonunitary states, as it vanishes after summation over spin for unitary pairings. The numerator of this term can be written down in full as,
\begin{equation}
\left(\epsilon_{a\bk} - \epsilon_{b\bk} +\frac{|\Delta^{ss}_{aa,\bk}|^2}{2\epsilon_{a\bk}}-\frac{|\Delta^{ss}_{bb,\bk}|^2}{2\epsilon_{b\bk}} + \frac{|\Delta^{ss}_{ab,\bk}|^2}{2\epsilon_{b\bk}} - \frac{|\Delta^{ss}_{ab,\bk}|^2}{2\epsilon_{a\bk}} \right) \bV^{ab, ss}_{\bk}\times \bV_{\bk}^{ba,ss} \,.
\end{equation}
Substituting into Eq.~\eqref{eq:AHCeffectiveTheory2}, it is easy to see that intraband and interband pairings both contribute to the full Hall conductivity. In particular, interband pairing is not absolutely required, because the $\bk$ and spin summation can return finite value even when only nonunitary intraband pairing is present. This agrees with the conclusion obtained in the maintext.

\section{Model with low-frequency Hall resonances}
\label{app:C}
In this appendix we present an explicit calculation of a nonunitary superconducting model that exhibits Hall resonances at low frequencies near the superconducting gap edge. We consider a square lattice with $D_{4h}$ symmetry, comprising $d_{xz}$ and $d_{yz}$ orbitals on each lattice site. In the basis $\Psi_{\bk}^{\dagger}=(c_{xz,\bk\uparrow}^{\dagger},c_{xz,\bk\downarrow}^{\dagger},c_{yz,\bk\uparrow}^{\dagger},c_{yz,\bk\downarrow}^{\dagger})$, introducing Pauli matrices $\tau_{i}$ and $\sigma_{i}$ ($i=0,x,y,z$) that act, respectively, in orbital and spin space, the symmetry allows the following general normal-state Hamiltonian $H=\Psi^\dagger_{\bk}\hat{H}_{0\bk}\Psi_{\bk}$,
\begin{equation}
\hat{H}_{0\bk} = 
\epsilon_{xz, \bk}\tau_{+} +\epsilon_{yz,\bk}\tau_{-}
+ \xi_{\bk}\tau_{x} + \eta^{\text{soc}}_{\bk} \,,
\end{equation}
with $\tau_{\pm}=\frac{\tau_{0}\pm\tau_{z}}{2}$. Here $\epsilon_{xz,\bk} = -2t\cos k_{x}-2\tilde{t}\cos k_{y}-\mu_{x}$ and $\epsilon_{yz, \bk} = -2\tilde{t}\cos k_{x}-2t\cos k_{y}-\mu_{y}$ are the dispersion of the $d_{xz}$ and $d_{yx}$ orbitals, respectively; $\lambda_{\bk}=\lambda\sin k_{x}\sin k_{y}$ is the inter‑orbital hybridization; and $\eta^{\text{soc}}_{\bk}=\eta_{0}\tau_{y}\sigma_{z}+\eta^{\prime}(\sin k_{x}\sin k_{z}\tau_{y}\sigma_{x}+\sin k_{y}\sin k_{z}\tau_{y}\sigma_{y})$ describes the spin‑orbit coupling term.

Using the Nambu spinor $\Phi_{\bk}^{\dagger} = \bigl( \Psi_{\bk}^{\dagger},\; \Psi_{-\bk}^{T} \bigr)$, the BdG Hamiltonian matrix follows as
\begin{equation}
\hat{H}^{\mathrm{sc}}_{\bk} = 
\begin{pmatrix}
\hat{H}_{0\bk} & \hat{\Delta}_{\bk} \\
\hat{\Delta}_{\bk}^{\dagger} & -\hat{H}_{0\bar{\bk}}^{*}
\end{pmatrix}.
\end{equation}
We consider the following nonunitary $A_{1u}+iA_{2u}$ state constructed on the $d_{xz}$- and $d_{yz}$-orbital basis~\cite{ZhangJL2024PRL},
\begin{equation}
    \Delta_{\bk}(-i\sigma_{y})=\Delta_{A_{1u}} \left(\sin k_{x} \tau_{+}\sigma_{x}+  \sin k_{y} \tau_{-}\sigma_{y}\right)+i\Delta_{A_{2u}} \left(\sin k_{x} \tau_{+}\sigma_{y}-  \sin k_{y} \tau_{-}\sigma_{x}\right) \,,
\end{equation}

\begin{figure}
\includegraphics[width=8cm]{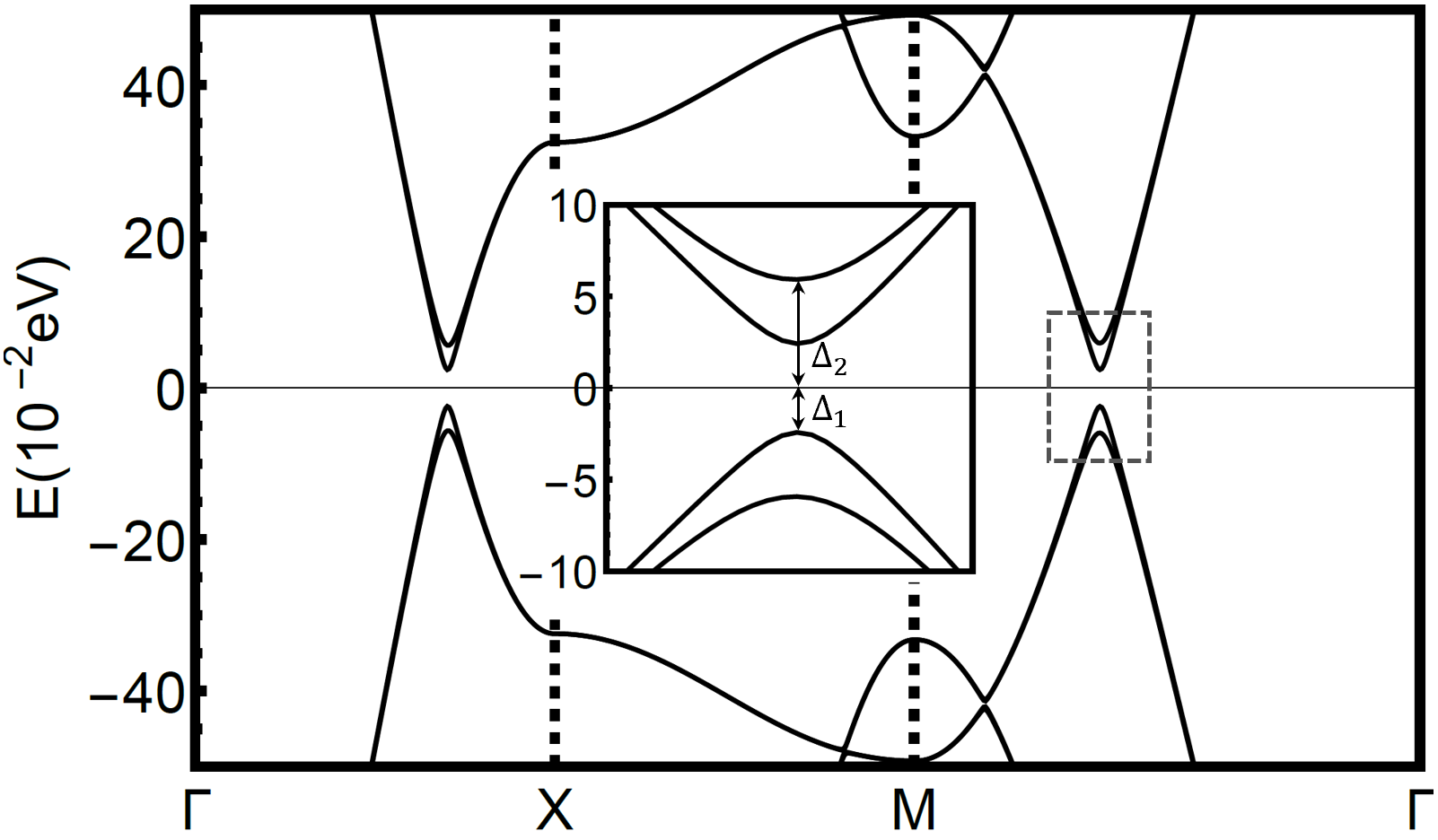}
\caption{The Bogoliubov band structure of the nonunitary $A_{1u}+iA_{2u}$ state along high symmetry path of Brillouin zone for the model introduced in App.~\ref{app:C}. We used the following set of parameters: $\left(t, \tilde{t},\lambda, \eta_{0},\eta^{\prime} \mu, \tilde{\mu}\right)=(0.4,0.04,0.08,0.1,0.1,0.4,1.2)~\mathrm{eV}$ with pairing amplitudes $\left( \Delta_{A_{1u}}, \Delta_{A_{2u}} \right)= (0.05,0.02 )~\mathrm{eV}$. For the unitary chiral $p$-wave model we used $\Delta_{p}=0.05 \mathrm{eV}$ (the corresponding Bogoliubov band structure is not shown). The inset provides an enlarged view of region highlighted by the gray dashed box. The smallest gap splitting are marked with $\Delta_{1}$ and $\Delta_{2}$. }
\label{fig:bandstr}
\end{figure}

The corresponding Bogoliubov band structure under a representative set of parameters is presented in Fig.~\ref{fig:bandstr}. We evaluate the optical Hall conductivity of this model following the standard linear-response theory given in the maintext. The result is shown in Fig.~3 in the maintext, where a low-frequency optical Hall resonance is clearly seen near $\omega = \Delta_1+\Delta_2$. Here, $\Delta_1$ and $\Delta_2$ stand for the gap size of the two positive Bogoliubov branches. Note that, while we have constructed the model in the orbital basis, qualitatively the same result can be obtained if the model is formulated in the band basis wherein the normal-state Hamiltonian is diagonalized. 

For comparison, we also evaluate the Hall conductivity of a unitary chiral $p$-wave state whose pairing matrix is given by
\begin{equation}
    \Delta_{\bk}(-i\sigma_{y})=\Delta_{p} \left(\sin k_{x} \tau_{+}\sigma_{z} +i\sin k_{x} \tau_{-}\sigma_{z} \right)  \,.
\end{equation}
The corresponding result is also shown in Fig.~3 in the maintext, showing no low-frequency resonance---consistent with our theory. 

%However, it was recently pointed out that, unless certain symmetry requirements are met, the transitions between the low-lying quasiparticle bands separated by $2\Delta$ are still prohibited. Accordingly, longitudinal optical conductivity at the frequency scale of $\omega \sim 2\Delta$

% Note that point group $C_{i}$ does not support non-unitary pairing, though it is inversion symmetric.

% But the absence of these symmetries cannot be equated with the presence of Hall effect. 

%  In multiband superconductors without interband Cooper pairing, the Bogoliubov quasiparticle solutions associated with different Bloch bands are fully decoupled. However, the velocity operators in general do not commute with the Hamiltonian. Therefore, interband velocity may couple quasiparticles from different bands, thereby enabling interband transitions. In other words, the aforementioned transition matrix element may not vanish if the two quasiparticle states originate from different Bloch bands. A pair of interband transition processes induced by interband velocity is schematically depicted in Fig.~\ref{fig:transitions}. 

% In this study, we focus on the Hall response in the $\bq=0$ limit. 

% In unitary TRSB superconductors, particle-hole and inversion symmetries prohibit optical Hall transitions at the superconducting gap energy scale. But low-frequency Hall transitions are allowed in nonunitary superconducting states.  

\end{widetext}

\end{document}